\documentclass[final,3p,times,authoryear]{elsarticle}

\usepackage[hidelinks]{hyperref}
\usepackage{amssymb}

\usepackage{amsmath,amsfonts}
\usepackage{algorithmic}
\usepackage{array}
\usepackage{textcomp}
\usepackage{stfloats}
\usepackage{url}
\usepackage{verbatim}
\usepackage{graphicx}
\usepackage{multirow}
\usepackage{subfigure}  
\usepackage{makecell}
\usepackage{longtable}
\usepackage[hidelinks]{hyperref}
\usepackage{subfigure}  
\usepackage{makecell}
\usepackage{subcaption}
\usepackage{subfigure}  
\usepackage[ruled,linesnumbered]{algorithm2e}
\usepackage{makecell}
\usepackage{mathrsfs}
\usepackage{xcolor}
\usepackage{enumitem}
\usepackage[linesnumbered,ruled]{algorithm2e}

\begin{document}
\begin{frontmatter}

\title{The Swarm Intelligence Freeway-Urban Trajectories (SWIFTraj) Dataset - Part II: A Graph-Based Approach for Trajectory Connection}

\author[inst1]{Xinkai Ji}

\author[inst1]{Pan Liu}

\author[inst3,inst4]{Ying Yang}

\author[inst2]{Yu Han}
\ead{yuhan2@hkust-gz.edu.cn}


\affiliation[inst1]{organization={School of Transportation},
            addressline={Southeast University}, 
            city={Nanjing},
            country={China}}

\affiliation[inst2]{organization={Thrust of Intelligent Transportation},
            addressline={The Hong Kong University of Science and Technology (Guangzhou)}, 
            city={Guangzhou},
            country={China}}

\affiliation[inst3]{organization={School of Management},
            addressline={Shanghai University}, 
            city={Shanghai},
            country={China}}

\affiliation[inst4]{organization={Department of Architecture and Civil Engineering},
            addressline={Chalmers University of Technology}, 
            city={Gothenburg},
            country={Sweden}}






\begin{abstract}

In Part I of this companion paper series, we introduced SWIFTraj, a new open-source vehicle trajectory dataset collected using an unmanned aerial vehicle (UAV) swarm. The dataset has two distinctive features. First, by connecting trajectories across consecutive UAV videos, it provides long-distance continuous trajectories, with the longest exceeding 4.5 km. Second, it covers an integrated traffic network consisting of both freeways and their connected urban roads. Obtaining such long-distance continuous trajectories from a UAV swarm is challenging, due to the need for accurate time alignment across multiple videos and the irregular spatial distribution of UAVs. To address these challenges, this paper proposes a novel graph-based approach for connecting vehicle trajectories captured by a UAV swarm. An undirected graph is constructed to represent flexible UAV layouts, and an automatic time alignment method based on trajectory matching cost minimization is developed to estimate optimal time offsets across videos. To associate trajectories of the same vehicle observed in different videos, a vehicle matching table is established using the Hungarian algorithm. The proposed approach is evaluated using both simulated and real-world data. Results from real-world experiments show that the time alignment error is within three video frames, corresponding to approximately 0.1 s, and that the vehicle matching achieves a consistently high F1-score. These results demonstrate the effectiveness of the proposed method in addressing key challenges in UAV-based trajectory connection and highlight its potential for large-scale vehicle trajectory collection.

\end{abstract}


\end{frontmatter}

\section{Introduction}

High-resolution trajectory data is invaluable in road traffic flow research. It has been extensively used for understanding driving behavior, analyzing the mechanisms behind puzzling traffic phenomena, calibrating (training) and validating traffic flow models, and optimizing traffic flow dynamics. In recent years, there is an increasing availability of open-source trajectory datasets, further advancing the research in this regard. In particular, unmanned aerial vehicles (UAVs), combined with advanced computer vision techniques, have enabled the extraction of high-resolution vehicle trajectories from aerial videos \citep{chen2020high,liu2021vehicle,shi2021video}.

Despite these advances, most existing UAV-based trajectory datasets remain limited in spatial coverage. For instance, the trajectory data on US-101 in the NGSIM dataset \citep{alexiadis2004next}, captured from high buildings, cover a maximum distance of 640 meters. The highD dataset \citep{krajewski2018highd}, recorded by a single UAV along German freeways, spans only about 420~m. Such limited spatial extent of vehicle trajectories hinders comprehensive observation of the full life cycle of traffic phenomena, such as capacity drop and oscillations \citep{chen2014periodicity}. While datasets such as I24-Motion \citep{gloudemans202324} cover a much broader spatial range, the trajectories are discontinuous, which limits their use for driving behavior analysis. Therefore, long-distance continuous trajectory datasets remain scarce in the transportation field.

To address this limitation, recent studies have explored the use of multiple UAVs to collect trajectory data over extended roadway segments \citep{barmpounakis2020new,ma2022magic}. A key challenge in this setting is connecting vehicle trajectories across multiple UAV videos. Existing approaches can be broadly categorized into two types. The first type stitches multiple videos into one before extracting trajectories \citep{chen2020modeling,chen2022empirical}. While effective for small-scale scenarios (typically two videos), this strategy typically involves a high computational cost and becomes impractical as the number of videos increases. The second type first extracts trajectories independently for each video and then connects trajectories across videos \citep{shi2021car,raju2022developing,coifman2024partial}. This approach significantly reduces computational complexity, making it applicable to large-scale trajectory connection tasks. Hence, this paper focuses on this type of approach.

Nevertheless, most existing methods of the second type face the following limitations. First, time alignment across videos is often neglected, which may lead to vehicle matching errors and discontinuities in the connected trajectories. Second, many existing methods focus on connecting trajectories within the same lane and have limited capability to handle lane-changing scenarios. Furthermore, current connection methods \citep{shi2021car,coifman2024partial,wang2024automatic} are primarily designed for scenarios where multiple UAVs are arranged in a line, as illustrated in Figure~\ref{fig:video_graph_m:b}. As a result, these methods are unable to handle situations where UAVs are distributed in a scattered manner, as depicted in Figure~\ref{fig:video_graph_p:a}. The irregular distribution of UAVs implies that vehicle trajectories in one video may need to be connected with trajectories from three or four surrounding videos, which significantly increases the complexity of trajectory connection. In summary, the challenges of large-scale UAV-based trajectory connection can be attributed to two key aspects: time alignment across multiple videos and the irregular distribution of UAVs.




To address these challenges, this paper proposes a Graph-based approach for Connecting Vehicle Trajectories from a UAV swarm (GCVT), which consists of two subtasks: time alignment and vehicle re-identification (Re-ID). Time alignment, also known as time synchronization, aligns the timestamps of vehicle trajectories captured by different UAVs. Vehicle Re-ID refers to the task of identifying the same vehicle across multiple cameras \citep{li2023tvg,wang2023blockchain,sarker2024transformer}. In computer vision, vehicle Re-ID typically relies on image features for matching. However, due to the high flying altitude of UAVs and the resulting bird’s-eye-view perspective, vehicles in the imagery are often small, making image-based matching challenging. To overcome this limitation, we adopt a trajectory-based matching strategy instead of relying on image features. Specifically, vehicles are matched based on their trajectories across adjacent videos. The matching relationships are established between vehicles in neighboring videos, and the associations are then extended to all videos. This process ensures that each vehicle maintains a consistent ID across the entire dataset.

To this end, the proposed GCVT comprises the following key steps. First, a graph is constructed to represent the UAV layout, with each video treated as a node and edges defined based on overlapping views, allowing the approach to handle irregular UAV distributions. Next, an automatic time alignment method based on trajectory matching cost minimization is applied to estimate pairwise time offsets, which are then accumulated to achieve global time synchronization. Finally, vehicle trajectories across adjacent videos are associated using the Hungarian algorithm \citep{kuhn1955hungarian}, and continuous trajectories are generated through trajectory fusion. The effectiveness of the proposed approach is demonstrated through both simulation experiments and real-world UAV swarm experiments. Overall, the main contributions of this study are summarized as follows.


(1) A graph-based approach for connecting vehicle trajectories captured by a UAV swarm is proposed, enabling the generation of continuous trajectories across multiple UAV videos under irregular spatial distributions. 


(2) The time offset optimal algorithm based on trajectory matching cost is proposed for the automatic time alignment of multiple videos. This ensures the temporal synchronization of the trajectories in these videos.

(3) The effectiveness of the proposed approach is evaluated through a real-world experiment conducted with a swarm of 16 UAVs. This evaluation assesses the accuracy of the time alignment and vehicle matching. The results demonstrate the effectiveness of the GCVT approach in real-world scenarios. Meanwhile, the vehicle trajectory dataset from 16 UAVs is available online for non-commercial research purposes.

The rest of this paper is organized as follows. The next section introduces the related work. Section~\ref{sec:problem} presents some key definitions and the task of vehicle trajectory connection. Section~\ref{sec:meth} thoroughly expounds upon the GCVT approach, explaining the step-by-step process of connecting the vehicle trajectories. Section~\ref{sec:exp} reports the experimental setup and evaluates the performance of time alignment and vehicle matching. Finally, Section~\ref{sec:conclusion} concludes the paper and discusses directions for future research.

\section{Related Work}
\label{sec:related_work}
\subsection{Trajectory Dataset}
Methods for obtaining vehicle trajectories can be divided into two categories: vehicle-based methods and video-based methods \citep{kim2010evaluation}. The vehicle-based methods rely on onboard sensors such as Lidar, Camera, GPS, and IMU to gather trajectories of the ego vehicle and surrounding traffic participants \citep{coifman2016collecting,zhao2016road,hu2022processing,chen2023ecmd,LIU2025100200}. These methods are capable of capturing long-duration and long-distance trajectories. However, the high cost of sensor-equipped vehicles and the low penetration rate of such vehicles limit their applicability for large-scale traffic flow analysis. Video-based methods extract vehicle trajectories from traffic videos \citep{xu2016enhanced,chen2020high,liu2021vehicle,shi2021video}. Compared with vehicle-based approaches, video-based methods are more cost-effective and can provide full-sample trajectory data. The emergence of UAV technology has further sparked research interest in extracting vehicle trajectories from UAV videos. UAV-based data collection offers several advantages, including a bird’s-eye view for comprehensive observation of traffic dynamics, high flexibility in capturing diverse traffic scenarios, and lower deployment cost relative to vehicle-based sensing.

In recent years, UAVs have become the primary method for obtaining vehicle trajectories. Krajewski et al. released a series of UAV-based trajectory datasets, including HighD \citep{krajewski2018highd}, exiD \citep{moers2022exid}, inD \citep{bock2020ind}, and roundD \citep{krajewski2020round} datasets, each focusing on different roadway sections such as basic highway sections, ramps, intersections, and roundabouts. Additionally, the INTERACTION dataset \citep{zhan2019interaction} is specifically designed for driver behavior-related research and comprises trajectory data in four scenarios: roundabouts, un-signalized intersections, signalized intersections, and merging. To analyze the driving behavior of Chinese drivers and the characteristics of Chinese roads, many scholars have published vehicle trajectory data collected in China. The SIND dataset \citep{xu2022drone} provides vehicle trajectories within two-phase intersections in Tianjin, China. \cite{chen2020high} published vehicle trajectories on various urban expressway segments during both peak and non-peak hours on weekdays in Nanjing, China. \cite{zhang2023ad4che} focused on traffic congestion scenarios and proposed vehicle trajectory data for China's congested highways, aiming to develop traffic jam pilot systems. Despite their value, these datasets are primarily collected by a single UAV, which inherently limits the spatial extent of the resulting vehicle trajectories.

To obtain longer vehicle trajectories, some studies have adopted multi-UAV data collection schemes. The pNEUMA experiment \citep{barmpounakis2020new} deployed a swarm of ten drones to capture a large-scale urban traffic dataset in the central business district of Athens, Greece. This effort provided a high-resolution, multi-modal traffic dataset, enabling in-depth analysis of congestion phenomena and offering a valuable resource for global researchers to develop and test traffic models. \cite{ma2022magic} proposed the MAGIC dataset, which was collected on an expressway in Shanghai, China, using 6 UAVs. The dataset covers a total road length of 4,000 meters in both directions.
However, in these studies, the processing of UAV videos relies on commercial platforms, such as DataFromSky \citep{datafromsky2023}. The underlying processing pipelines and algorithms are not publicly disclosed, which limits the reproducibility and transparency of the resulting datasets. In addition, the datasets produced in these studies typically cover a limited range of road types.

\subsection{Trajectory Connection}
Methods for obtaining continuous vehicle trajectories from multiple UAV videos can generally be categorized into two types: early connection \citep{chen2020modeling,chen2022empirical,zheng2024citysim,Chaudhari2025MiTra,Rajput2026SPT}, in which videos are first stitched together, and late connection \citep{shi2021car,raju2022developing,coifman2024partial}, in which trajectories extracted from individual videos are subsequently connected to form complete vehicle trajectories.

In early connection approaches, multiple UAV videos are aligned and stitched into a single composite video, after which conventional single-video trajectory extraction methods are applied \citep{chen2020modeling,chen2022empirical,BERGHAUS2024100133}. This process typically relies on affine transformations to align and merge video frames into a unified video stream. While effective for scenarios involving a small number of videos, early connection methods are generally limited to stitching only two or three videos, as the computational cost increases rapidly with the number of videos. Additionally, the process of video stitching faces the significant challenge of video time alignment, also known as video time synchronization. Since different UAVs may have variations in their internal clock systems, the timestamps carried by each video may exhibit temporal discrepancies. Therefore, manual time alignment is necessary to ensure accurate synchronization of all the videos. However, as the number of videos increases, the workload for manual time alignment will significantly increase. As the number of videos increases, the manual effort required for time alignment grows substantially, making the process increasingly complex, time-consuming, and resource-intensive. 

The late connection strategy extracts vehicle trajectories independently from each video and subsequently connects them across videos. Compared with early connection methods, this strategy offers several advantages. First, this approach demands lower computing resources, making it more feasible and practical for processing a larger number of videos. Second, trajectory extraction can be performed independently for each video, which enables parallel processing and significantly accelerates the overall pipeline. For these reasons, the late connection strategy is adopted in this study. 

In the late connection method, a car-following-based vehicle trajectory connection method was proposed to connect broken vehicle trajectories \citep{shi2021car,shi2021video}. This method formulates trajectory connection across multiple videos as a problem analogous to trajectory interruptions caused by bridge occlusions. However, bridge-induced occlusions occur within a single video and do not involve time misalignment across different videos. Additionally, the method encounters challenges when attempting to connect vehicle trajectories in situations where vehicles change lanes within the overlapping area of two videos, mainly due to the utilization of the car-following model. 
\cite{raju2022developing} proposed a stitching-based algorithm to obtain continuous long trajectories. However, this approach solely relies on the two endpoint data of the trajectory, leading to poor reliability in vehicle matching.
\cite{coifman2024partial}  introduced a partial trajectory method to align views from successive fixed cameras, facilitating video-based vehicle tracking across multiple camera views. However, these methods do not consider the time synchronization between multiple videos.
When significant time offsets exist between trajectories captured by adjacent videos, achieving a robust trajectory connection becomes challenging. 
Therefore, to overcome the limitations of existing late connection methods and meet the requirements of large-scale, multi-UAV trajectory data collection, a new trajectory connection approach that explicitly accounts for time alignment and complex deployment scenarios is required.

\section{Problem statement}
\label{sec:problem}
The problem of connecting vehicle trajectories collected by multiple UAVs focuses on generating a unified and continuous trajectory representation from multiple UAV videos. 
In the proposed GCVT framework, vehicle trajectories extracted from individual videos constitute the primary input data. It is worth noting that video stabilization and trajectory extraction are not the focus of this paper. For specific methods, please refer to our research \citep{ji2024openvter}.
For successful vehicle trajectory connection, two prerequisites need to be fulfilled.
(1) All vehicle trajectories have the same time interval. (2) Any pair of videos to be connected must have overlapping image regions, within which vehicle trajectories are observable.
Moreover, we assume that all videos are recorded at a frame rate of $\frac{1}{\Delta t}$, where ${\Delta t}$ represents the time interval between consecutive frames. The vehicle trajectories are extracted at a time interval of $A{\Delta t}$, where $A$ is a constant factor determining the sampling rate of the trajectory. In this study, $\Delta t$ is set to $1/30~\mathrm{s}$ and $A=3$.
Based on these assumptions, we subsequently formulate the problem using a graph-theoretical framework to enable a rigorous mathematical representation. 
Next, some key definitions and the task are stated as follows.

\textbf{Definition 1: Video Graph.} 
The spatial relationship among videos captured from different locations is defined as a video graph. Specifically, an undirected graph $G=(V,E)$ is constructed to represent the topological structure of the video layout. Each video is treated as a node, and the node set is defined as $V=\{v_m \mid m=1,\ldots,M\}$, where $M$ denotes the total number of videos. The edge set is defined as $E=\{e_{i,j} \mid i,j \in \{1,\ldots,M\},\, i \neq j\}$, where an edge $e_{i,j}$ exists if and only if videos $v_i$ and $v_j$ have overlapping image regions.

\begin{figure}[!ht]
	\centering
	\subfigure[]{\includegraphics[width=0.8\textwidth] {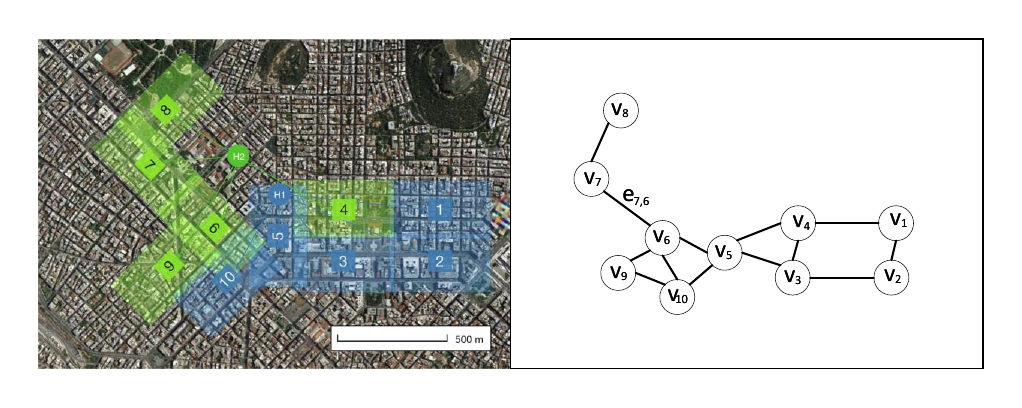}\label{fig:video_graph_p:a}}
	\subfigure[]{\includegraphics[width=0.8\textwidth] {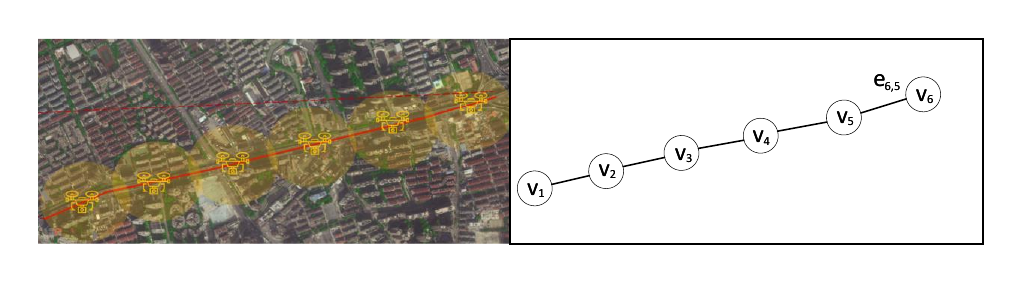} \label{fig:video_graph_m:b}}
	\caption{Example of video graph. (a) (Left) The flight plan of pNEUMA dataset (the image is from \cite{barmpounakis2020new}). (Right) The video graph of pNEUMA dataset. (b) (Left) The flight plan of MAGIC dataset (the image is from \cite{ma2022magic}). (Right) The video graph of MAGIC dataset.}\label{fig:video_graph}
\end{figure}

To illustrate the structure of the video graph, the pNEUMA dataset \citep{barmpounakis2020new} and MAGIC dataset \citep{ma2022magic} are taken as representative examples, and their corresponding video graphs are illustrated in Fig.~\ref{fig:video_graph}.
Taking Fig.~\ref{fig:video_graph_p:a} as an example, the construction of the video graph is described as follows. First, the video recorded at each collection position is represented as a node. For example, there are 10 collection positions in Fig.~\ref{fig:video_graph_p:a}, and the corresponding video nodes are $v_1-v_{10}$. Then, an edge is established between two nodes if their corresponding videos contain overlapping image regions. For example, since there is overlap between $v_6$ and $v_7$, edge $e_{7,6}$ is included in the graph.

\textbf{Definition 2: Video Timestamp.} The starting timestamp of video $v_m$ is denoted as $T_m$. Due to the difference between the UAV onboard clock and the real-world time, timestamps across different videos may not be temporally synchronized. The real starting timestamp is denoted as $\hat{T}_m = T_m + \varepsilon_m$, where  $\varepsilon_m$ is the unknown time error of video $v_m$. In practice, the time error is not directly available. Instead, the measurable quantity is the pairwise time offset between two videos. For an edge connecting videos $v_i$ and $v_j$, the time offset is defined as $\delta_{i,j}= \varepsilon_i - \varepsilon_j$. When $\delta_{i,j}=0$, the two videos are temporally aligned.

\textbf{Definition 3: Main Video and Sub Video.} 
In the vehicle trajectory connection problem formulated on the undirected graph $G=(V,E)$, a reference node is designated to provide a unified temporal and spatial reference across all videos. We define the reference node as the main video $v_a$ ($a \in [1,...,M]$). All remaining nodes in the graph are referred to as sub videos.
As the temporal reference, the timestamp of the main video is considered to represent the real-world time, i.e., its time error is assumed to be zero ($\varepsilon_a = 0$). As the spatial reference, the pixel coordinate system of the main video is utilized as the coordinate system for the final connected trajectories. 
 
\textbf{Definition 4: Vehicle Trajectory.} A vehicle trajectory is defined as a time-ordered sequence of spatial positions describing a vehicle over time. 
Let $C^m = \{c_i^m \mid i = 1,\ldots,N_m\}$ denote the set of vehicles observed in video $v_m$, where $N_m$ is the total number of vehicles in $v_m$. The trajectory of vehicle $c_i^m$ is represented as
$J_i^m = \left\{ \left( \boldsymbol{p}_k^i ,\, l_k^i \right) \;\middle|\; k = 1,\ldots,K \right\},$
where $\boldsymbol{p}_k^i = (x_k^i, y_k^i)$ denotes the pixel coordinates of the vehicle center at the $k$-th sampled trajectory frame, and $l_k^i$ denotes the corresponding lane ID at frame $k$. Trajectories are extracted at a sampling interval of $A\Delta t$, such that the $k$-th trajectory frame corresponds to the $\big(A(k-1)+1\big)$-th frame of the original video.



\textbf{Definition 5: Vehicle Matching Table.} 
For each edge $e_{i,j} \in E$, the objective of trajectory connection is to establish the correspondence between vehicle trajectories observed in videos $v_i$ and $v_j$. The correspondence is represented by a vehicle matching table $R_{i,j}$.
Specifically, $R_{i,j}$ is defined as a set of matched vehicle pairs, which consists of three columns: vehicle ID in video $v_i$, vehicle ID in video $v_j$, and the trajectory matching cost. The trajectory matching cost quantifies the dissimilarity between two vehicle trajectories and is formally defined in Section~\ref{sec:meth}.


\textbf{Task Formulation.} Based on the above definitions, the vehicle trajectory connection problem is formulated as follows. \textbf{Input:} the vehicle trajectories in all videos $J = \left\{ J_i^m \mid m = 1,\ldots,M,\; i = 1,\ldots,N_m \right\}$, the starting timestamp of all videos $T = \{ T_m \mid m = 1,\ldots,M \}$, video graph $G=(V,E)$, and the selected main video $v_a$.
\textbf{Output:} the time error of all the videos $\epsilon = \{\epsilon_{m}\}$($m \in [1,...,M]$), 
the vehicle matching tables of edges ${R} = \{{R}_{i,j}\}$. Based on the estimated time errors and vehicle matching tables, individual trajectories are temporally aligned and fused to generate continuous vehicle trajectories.

\section{Methodology}
\label{sec:meth}

\subsection{Overview}

\begin{figure}[!ht]
	\centering
	\includegraphics[width=0.7\textwidth]{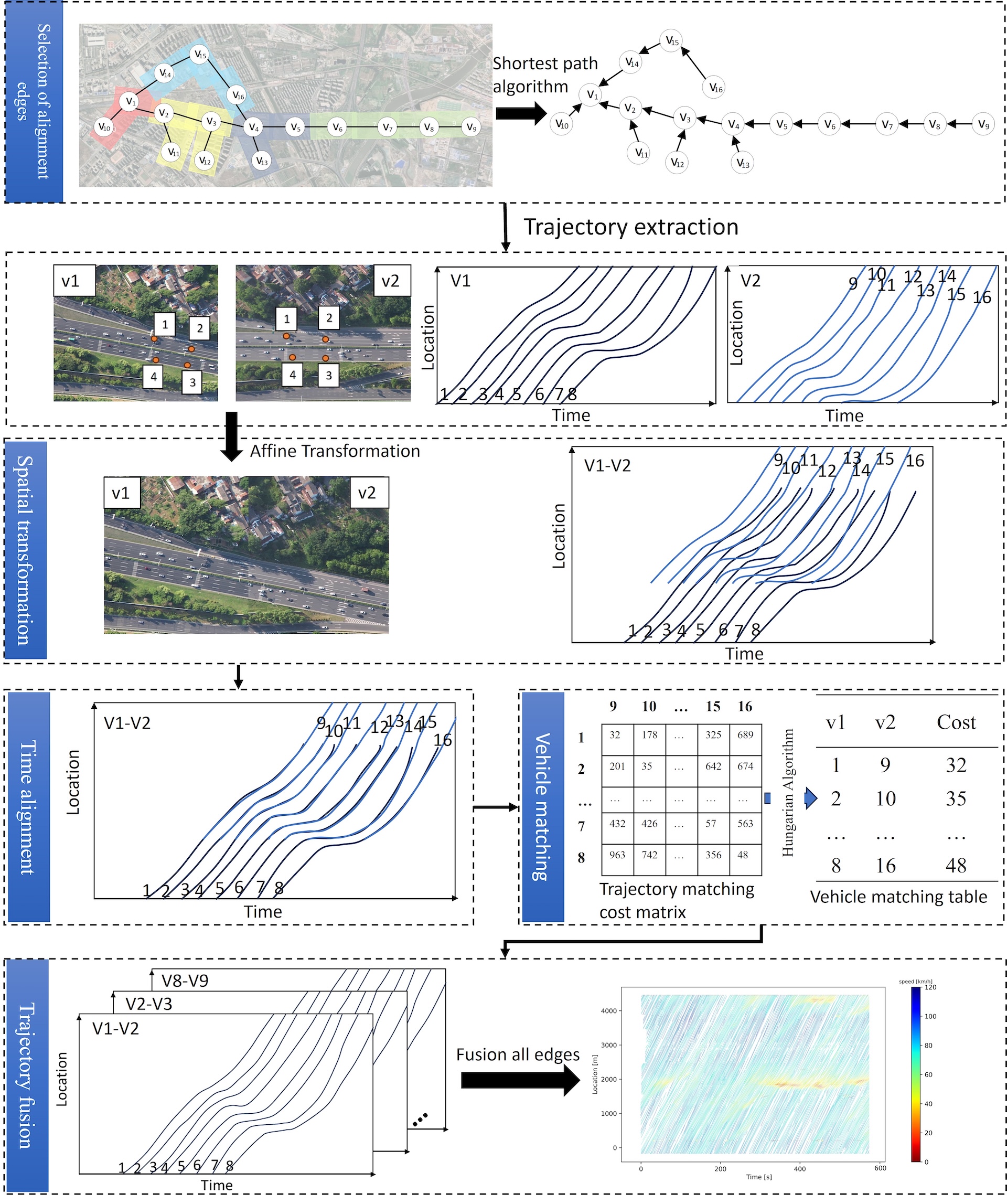}
	\caption{The GCVT framework. The subsets of trajectory data from video $v_1$ and $v_2$ are selected to demonstrate the trajectory connection process.}\label{fig:GCVT}
\end{figure}

The process of trajectory connection can be divided into five steps: selection of alignment edges, spatial transformation, time alignment, vehicle matching, and trajectory fusion. The overall framework of the GCVT approach is depicted in Fig.~\ref{fig:GCVT}, showcasing the integration of the five steps to realize trajectory connection from multiple UAV videos. The overall process is described as follows:

\textbf{Step 1:} Construct the alignment edge set $S_a$ based on the video graph.

\textbf{Step 2:} Estimate the affine transformation matrix for each edge in $S_{a}$ and compute the cumulative transformation matrix from each sub-video to the main video, denoted as $H_{i,a}$.

\textbf{Step 3:} Estimate the pairwise time offsets for edges in $S_{a}$ and derive the time error of each sub-video, denoted as $\epsilon_i$.

\textbf{Step 4:} Apply spatial transformation and time alignment to the vehicle trajectories in sub-videos based on $H_{i,a}$ and $\epsilon_i$, and subsequently calculate the corresponding vehicle matching table $R_{i,j}$ for each edge in $S_{a}$. 

\textbf{Step 5:} Connect vehicle trajectories based on the vehicle matching table to generate continuous vehicle trajectories.

Detailed formulations and algorithms for alignment edge construction, spatial transformation, time alignment, and vehicle matching are presented in the following subsections.

\subsection{Selection of alignment edges}
Errors are inevitable in both time alignment and spatial transformation between video pairs. 
As alignment parameters are propagated along longer paths from a sub-video to the main video, pairwise errors may accumulate, resulting in amplified global synchronization deviations.

To mitigate error accumulation, a shortest-path-based strategy is adopted to determine a compact and reliable subset of alignment edges. 
Here, alignment edges refer to the graph edges on which spatial transformation and time alignment are explicitly performed.  
Specifically, the video graph is treated as an unweighted directed graph with unit edge weights. For each sub-video, the shortest path to the main video is computed using a standard shortest-path algorithm (e.g., Dijkstra’s algorithm). All directed edges appearing along these shortest paths are collected and merged to form the alignment edge set $S_a$.
By restricting subsequent spatial and temporal alignment to this selected subset of edges, alignment parameters are propagated along shortest paths, effectively reducing cumulative errors and improving the robustness of the overall multi-video synchronization process.





\subsection{Spatial transformation}

\begin{figure}[!ht]
	\centering
	\includegraphics[width=0.48\textwidth]{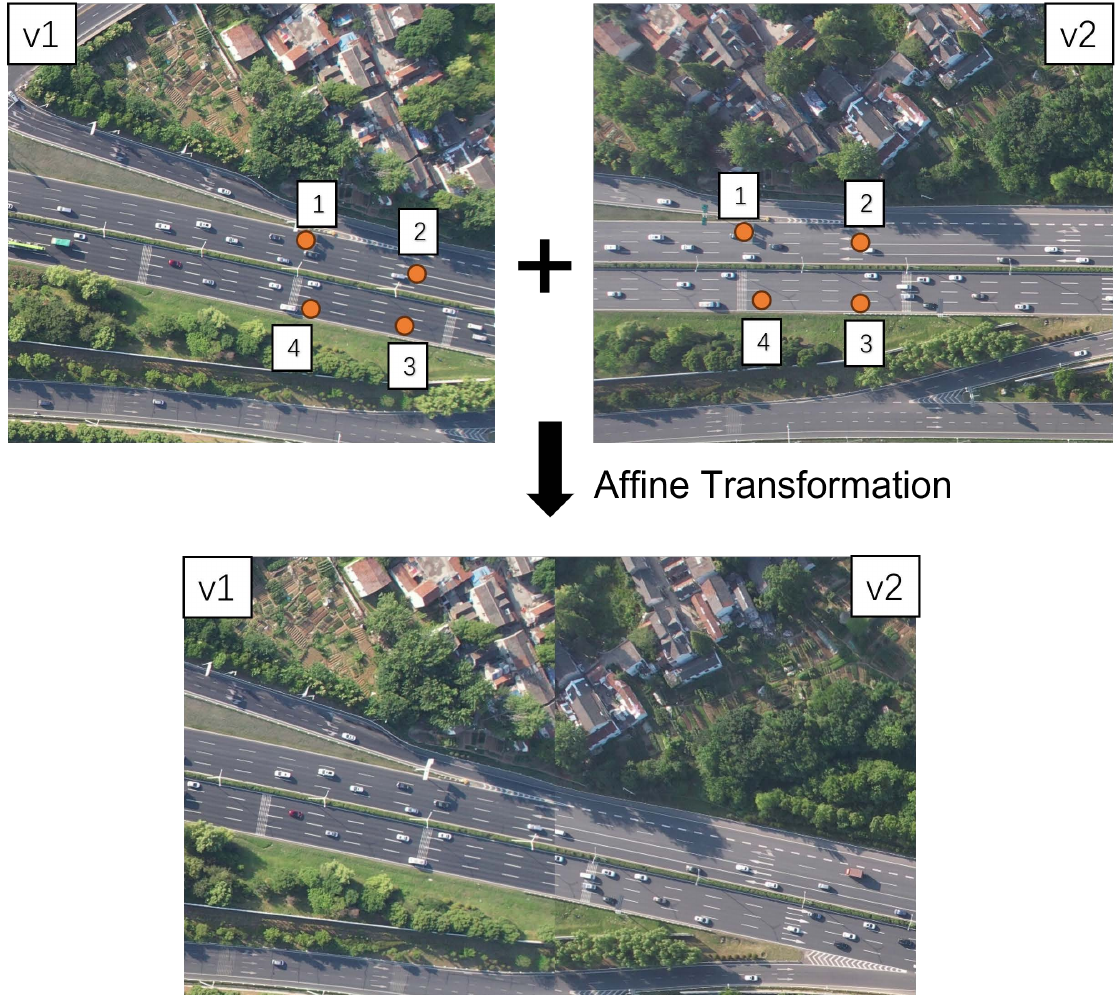}
	\caption{The spatial transformation of images. The four orange points are the selected feature points.}\label{fig:spatial_transformation}
\end{figure}

Accurate spatial transformation is critical for reliable trajectory connection. 
In UAV-based road scenes, automatic feature matching may become unstable due to non-coplanar structures, such as bridges or roadside objects. In such cases, background regions can be aligned while the roadway itself remains misaligned. To ensure the reliability of spatial transformation, a manual feature selection strategy is adopted in this study. Although the proposed approach involves manual annotation, it only requires annotating the first frame of each video, resulting in a manageable annotation workload. Specifically, for each alignment edge, a set of eight feature points (four points per video) is manually selected on the road surface within the overlapping region of the two videos, as illustrated in Fig.~\ref{fig:spatial_transformation}. For a directed edge $e_{i,j}$, the feature points selected in video $v_i$ are denoted by $\mathrm{pts}_{i,j}^{i}$. After annotating feature points for all alignment edges in $S_{a}$, the affine transformation matrices are estimated using the procedure described below.

The proposed spatial transformation algorithm consists of two main steps. First, for each space--time edge, the corresponding affine transformation matrix is estimated using the Random Sample Consensus (RANSAC) algorithm \citep{fischler1981random}, which provides robustness against potential annotation noise and outliers. Second, the affine transformation matrix from each sub-video to the main video is obtained by cumulatively composing the pairwise transformation matrices along the shortest path in the video graph. The detailed procedure is summarized in Algorithm~\ref{algorithm:spatial-transformation}.


\begin{algorithm}
\SetKwInOut{Input}{Input}
\SetKwInOut{Output}{Output}
\Input{Alignment edges $S_{a}$, feature point sets $\{{pts}_{i,j}^{i}\}$, video graph $G = (V, E)$, main video $v_a$ }
\Output{Affine transformation matrices from all sub-videos to the main video $H_{i,a}$}
Initialize the affine transformation matrix estimation function $\mathrm
{getAffine(\cdot)}$;\
\BlankLine
\ForEach{ edge $d_{i,j} \in S_{a}$}{
$H_{i,j} \leftarrow \mathrm{getAffine}(pts_{i,j}^i,pts_{i,j}^j)$;
}
\ForEach{$v_i \in V \setminus \{v_a\}$}{
    Compute the shortest path 
    $P_{i\to a} \leftarrow (d_{i,u_1}, d_{u_1,u_2}, \ldots, d_{u_m,a})$\;

    Obtain the affine transformation by cumulative composition:
    $H_{i,a} \leftarrow 
    H_{u_m,a} \cdots H_{u_1,u_2} H_{i,u_1}$\;
}
\caption{Spatial Transformation}
\label{algorithm:spatial-transformation}
\end{algorithm}

\subsection{Time alignment}
Time alignment is a critical component of the proposed GCVT framework, whose objective is to estimate the temporal offset of each sub-video relative to the main video. Accurate time offset estimation enables the synchronization of all UAV videos into a unified temporal reference.
To this end, we propose an automatic time alignment algorithm based on trajectory matching cost minimization. The algorithm is built on the principle that, when two videos are correctly time-aligned, vehicle trajectories within their overlapping spatial region exhibit the minimum trajectory matching cost. Owing to its reliance solely on trajectory data extracted from UAV videos, the proposed method is computationally efficient, resource-light, and broadly applicable. It can be readily extended beyond UAV swarms to other multi-camera traffic surveillance scenarios. Algorithm~\ref{algorithm:time-alignment} presents an efficient numerical solution to this optimization problem.

Specifically, the algorithm starts by utilizing the affine transformation matrix obtained in the spatial transformation step to extract vehicle trajectories within the overlapping region of the videos, which involves spatial filtering of the trajectories. Next, within a search window, it iterates through various frame offsets from 0 to find the optimal time offset. To reduce computational complexity, only vehicle trajectories within the overlapping temporal interval of the two videos—limited to the preceding 60 seconds—are considered. 
Let $n_i$ and $n_j$ denote the numbers of selected vehicle trajectories in videos $v_i$ and $v_j$, respectively. For a given pair of videos $v_i$ and $v_j$, consider trajectory $i$ extracted from $v_i$ and trajectory $j$ extracted from $v_j$. For a candidate time offset $\delta$, the trajectory matching cost is defined as



\begin{equation}
L_{i,j}(\delta)
=
\frac{1}{K}
\sum_{k=1}^{K}
\left[
\left\|
\boldsymbol{p}^{i}(t_k)
-
\boldsymbol{p}^{j}(t_k+\delta)
\right\|_2^2
+
\alpha\,\big|l^{i}(t_k)-l^{j}(t_k+\delta)\big|
\right],
\label{eq:trajectory_difference}
\end{equation}
where $t_k=(k-1)A\Delta t$, $\delta$ denotes the candidate time offset, $K$ is the number of frames, $\boldsymbol{p}=[x,y]$ is the pixel coordinates of the vehicle center, $l_k^{i}$ and $l_k^{j}$ represent the lane IDs of vehicle $i$ and $j$ at frame $k$, respectively. The parameter $\alpha$ controls the weight of the lane ID difference and is set to $100$ in this study. To avoid matching vehicles that move in opposite directions, a motion direction check is applied. Specifically, the displacement between the first and last sampled positions of each trajectory is used to approximate its motion direction. If $(\boldsymbol{p}_K^i - \boldsymbol{p}_1^i)\cdot(\boldsymbol{p}_K^j - \boldsymbol{p}_1^j) \le 0,$ 
the matching cost is set to $+\infty$ to avoid matching spatially proximate but oppositely moving vehicles. It is worth noting that the proposed trajectory matching cost is defined directly in the two-dimensional pixel coordinate space, rather than relying on longitudinal car-following relationships. This design enables the method to handle general traffic scenarios, including lane-changing and merging.



 
Pairwise trajectory matching costs are computed between all selected trajectories in $v_i$ and $v_j$, yielding a cost matrix of size $n_i \times n_j$. The Hungarian algorithm is then employed to obtain the optimal trajectory matching. Matched pairs with costs exceeding a threshold $\theta^{td}$ are discarded to eliminate unreliable associations. Let $\mathcal{M}$ denote the set of valid matched trajectory pairs under offset $\delta$ after threshold filtering. 
The optimal time offset is determined by minimizing the mean matching cost over the search window $w$,
\begin{equation}
\delta_{i,j}
=
\arg\min_{\delta}
L_{\mathrm{mean}}(\delta)
=
\arg\min_{\delta}
\frac{1}{|\mathcal{M}|}
\sum_{(i,j)\in\mathcal{M}}
L_{i,j}(\delta),
\end{equation}
To accelerate convergence, a bidirectional search strategy is adopted, starting from zero offset and exploring both positive and negative directions. An early stopping criterion is employed to terminate the search once the cost increases significantly relative to the historical minimum. 
Finally, the time error of each sub-video is obtained by cumulatively summing the pairwise time offsets along the shortest path from the sub-video to the main video in the video graph.

\begin{algorithm}
\SetKwInOut{Input}{Input}
\SetKwInOut{Output}{Output}
\Input{Alignment edges $S_{a}$, affine transformation matrices $H_{i,j}$, video graph $G = (V, E)$, main video $v_a$}
\Output{Time error of sub videos $\epsilon_i$}
Initialize search window $w$, $L_{\min}\!\leftarrow\!+\infty$, thresholds $\theta^{\min}$, $\theta^{td}$, scale factor $\theta^{scale}$, and sampling interval $A\Delta t$\;

\For{each $d_{ij} \in S_{a}$}{
Transform the first frame $F_j$ and trajectories $C^j$ of video $v_j$ based on the affine transformation matrix $H_{i,j}$, $\hat{F}_j \leftarrow H_{i,j} \cdot F_j $, $\hat{C}^j \leftarrow H_{i,j} \cdot C^j$;
\

Obtain the overlapping region $O_{i,j}$ between $v_{i}$ and $v_j$, $O_{i,j} \leftarrow F_i \cap \hat{F}_j$; \

Filter the trajectories $C^i$ and $\hat{C}^j$ within $O_{i,j}$ to get the new trajectories $C_{o}^i$ and $\hat{C_o}^j$;\

\For{$f$ = $0$ to $w$}{
\For{$d \in [-1,1]$}{
$\Delta f \leftarrow f*d$; $\delta_{i,j} \leftarrow \Delta f A\Delta t$;

Update the frame index of the trajectory $\hat{C_o}^j$, $\hat{k} \leftarrow \lceil k+(T_i-T_j)/(A\Delta t) + \Delta f \rceil$;\

Select trajectories $C_o^i$ and $\hat C_o^j$ within the overlapping temporal window of length 60s\;

Compute cost matrix $\mathbf{D}$ using Eq.~\eqref{eq:trajectory_difference}\;
Hungarian matching on $\mathbf{D}$; discard pairs with cost $>\theta^{td}$ to obtain $\mathcal{M}$\;
Compute $L_{\mathrm{mean}}(\delta)=\frac{1}{|\mathcal{M}|}\sum_{(i,j)\in\mathcal{M}}L_{i,j}(\delta)$\;

\If{$L_{\text{min}} > L_{\mathrm{mean}}(\delta)$}{
$L_{\text{min}} \leftarrow L_{\mathrm{mean}}(\delta)$;
$d_{min} \leftarrow d$; 
$\delta^*_{i,j} \leftarrow \delta_{i,j}$
}
\If{$d=d_{min}$ and $L_{\text{min}} < \theta^{min}$ and $L_{\mathrm{mean}}(\delta) > max(\theta^{scale}*L_{\text{min}},\theta^{min})$}{
Break out of two nested loops  \tcp{Early stopping strategy}
}
}
}

}

\For{$v_i \in V$ where $v_i \neq v_a$}{
    Compute the shortest path 
    $P \leftarrow (d_{i,u_1}, d_{u_1,u_2}, \ldots, d_{u_m,a})$\;

    $\epsilon_i \leftarrow 
    \delta^*_{i,u_1}+\delta^*_{u_1,u_2}+\cdots+\delta^*_{u_m,a}$\;
}

\caption{Time offset optimal algorithm}
\label{algorithm:time-alignment}
\end{algorithm}

\subsection{Vehicle matching}
\label{sec:vehicle_matching}
The purpose of vehicle matching is to obtain a vehicle matching table. In the time alignment step, a partial vehicle matching table has already been obtained. However, for computational efficiency, only vehicle trajectories within a 60-second window were selected for matching. To obtain the complete vehicle matching table, it becomes essential to calculate the trajectory matching cost matrix for all vehicle trajectories within the overlapping region after time alignment. We employ a two-step matching approach based on the Hungarian algorithm to improve the accuracy of the matching results and prioritize vehicle pairs with smaller trajectory matching cost. The specific steps are as follows:

\textit{Step 1}: Set the elements in the trajectory matching cost matrix $\mathbf{D}$ that have values greater than a threshold $\theta^{matching}$ to infinity, resulting in the new matrix $\mathbf{D}^{1}$. This is implemented to ensure that smaller trajectory matching costs are given higher priority in the matching process.

\textit{Step 2}: Apply the Hungarian algorithm to the new matrix $\mathbf{D}^{1}$  to obtain the first-step matching table $\mathbf{R}^{1}$.

\textit{Step 3}: Remove the already matched vehicles from the trajectory matching cost matrix $\mathbf{D}$, resulting in the updated matrix $\mathbf{D}^{2}$. Then, perform the second-step matching using the Hungarian algorithm on the updated matrix $\mathbf{D}^{2}$ to get the second-step matching table $\mathbf{R}^{2}$.

\textit{Step 4}: Combine $\mathbf{R}^{1}$ and $\mathbf{R}^{2}$ to obtain the final vehicle matching table $\mathbf{R}$. 


\section{Experiment}
\label{sec:exp}

In this section, we present two experiments conducted to evaluate the effectiveness of the proposed GCVT approach. These experiments include both numerical simulations and real-world applications, providing a comprehensive evaluation of the method.

\begin{figure}[!ht]
	\centering
	\includegraphics[width=0.65\textwidth]{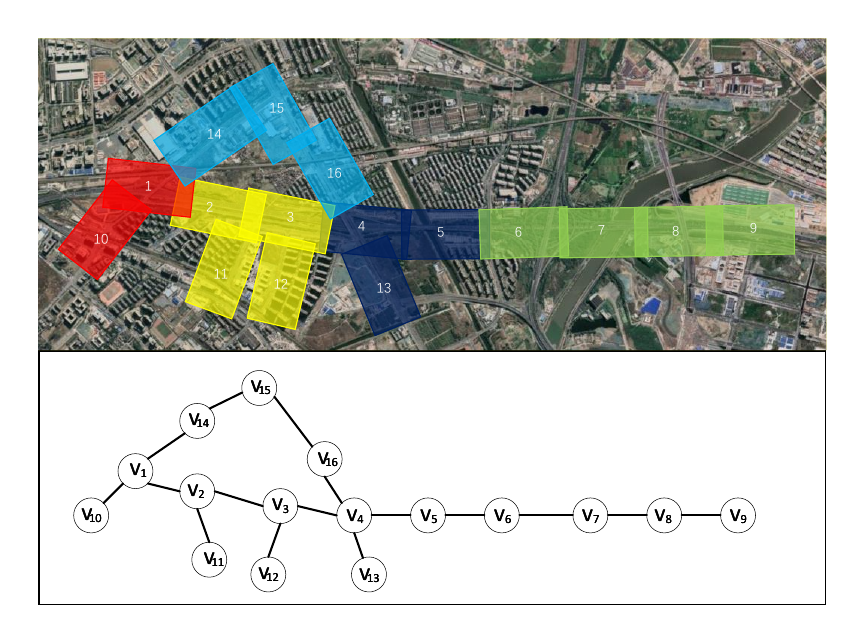}
	\caption{Study area and the video graph. (Up) Satellite map of the study area and the recording area of each UAV. (Down) Video graph for the UAV swarm}\label{fig:video_graph_h}
\end{figure}

\begin{itemize}
\item Experiment 1 (Exp1): This experiment uses the Swarm Intelligence Freeway-Urban Trajectories (SWIFTraj) dataset, collected by a swarm of 16 UAVs equipped with 5.4K-resolution cameras in a hybrid road environment comprising urban expressways, intersections, and multiple on- and off-ramps. This multi-UAV setup enables large-scale spatial coverage and provides the practical basis for graph construction, where each UAV video is treated as a node and overlap between neighboring videos defines the edges. The 5.4K-resolution cameras also allow each UAV to maintain sufficient vehicle observability at a relatively high flight altitude, which helps improve single-video trajectory quality and extend the effective coverage of each video for cross-video trajectory connection. Although these hardware settings are adopted in the present experiment, the framework itself is also applicable to other multi-UAV video data with overlap and reliable trajectories.

The objective is to evaluate the effectiveness of the proposed method under realistic traffic conditions.
Data were collected along a segment of the G42 Hurong Expressway and its adjacent intersections in Nanjing, China, as illustrated in Fig.~\ref{fig:video_graph_h}. Traffic flow was recorded during the morning peak hours (7:00–9:00 a.m.) on June 16 and 17, 2022. To ensure synchronized recording and reduce battery consumption during deployment, the UAVs were organized into five groups, with each group taking off from the same location.
As shown in Fig.~\ref{fig:video_graph_h}, UAVs with the same color belong to the same group. 

Video $v_1$ is designated as the main video. Videos $v_1$–$v_9$ were captured along the expressway mainline, while the remaining videos were recorded on adjacent urban roads. 
In the experiment, videos $v_1$–$v_9$ collected along the expressway mainline were primarily used to evaluate the reliability of the proposed method in constructing long-distance vehicle trajectories on urban expressways. In addition, four video nodes, including the expressway videos $v_{4}$ and $v_{5}$ and the urban-road videos $v_{13}$ and $v_{16}$, were selected  to evaluate the applicability of the proposed GCVT method under scenarios involving irregularly distributed UAVs.

\item Experiment 2 (Exp2): In this experiment, we used the microsimulation software, SUMO (Simulation of Urban MObility), to generate synthetic trajectory data as the ground truth. This simulation helped us collect vehicle trajectories across various regions, designated as the recording area of each UAV. The purpose of using synthetic trajectory data in this study is to analyze the accuracy of time alignment and vehicle matching under various time offsets and spatial transformation errors. By manually adjusting the time offsets and spatial transformation errors, we can test and evaluate the performance of the time alignment and vehicle matching under different conditions, ensuring they are robust and effective in real-world scenarios.

In the simulation, the road network and the coverage of each virtual UAV video collection point are designed to be consistent with the real conditions observed in Exp1. The duration of the simulation is set to 3600 seconds. The vehicle trajectories from two adjacent UAVs near the on-ramp area were selected to evaluate the time alignment and vehicle matching. Vehicles change IDs as they move across different UAV videos, necessitating vehicle matching. Gaussian noise with a mean $\mu$ and standard deviation $\sigma$ is added to the ground-truth position of the trajectory in one of the videos to test the accuracy of time alignment and vehicle matching under conditions of spatial transformation inaccuracies.


\end{itemize} 

The GCVT approach was developed in Python. The experimental platform of this research is a workstation with Intel Xeon W-2255 CPU@3.9GHz, and 64G memory.



\subsection{Evaluation metrics}
\label{sec:eval_metric}
Time alignment and vehicle matching are two critical components of the proposed trajectory connection framework, and their performance directly affects the accuracy and reliability of the final connected trajectories.

The performance of time alignment is evaluated using the time offset error, defined as the absolute difference between the estimated time offset $\delta_{\text{est}}$ and the ground-truth time offset $\delta_{\text{gt}}$: $|\delta_{\text{est}} - \delta_{\text{gt}}|$.


Vehicle matching performance is assessed using standard classification metrics, including Precision, Recall, and F1-score. These metrics are computed based on the following definitions:
\begin{itemize}
\item True Positive (TP): Number of correctly matched vehicle pairs.
\item False Positive (FP): Number of incorrectly matched vehicle pairs.
\item False Negative (FN): Number of ground-truth matches that were not identified.
\end{itemize} 

The evaluation metrics are defined as:
\begin{equation}
\text{Precision} = \frac{TP}{TP + FP},
\end{equation}
\begin{equation}
\text{Recall} = \frac{TP}{TP + FN},
\end{equation}
\begin{equation}
\text{F1-score} = 2 \cdot \frac{\text{Precision} \cdot \text{Recall}}{\text{Precision} + \text{Recall}}.
\end{equation}

Higher values of Precision, Recall, and F1-score indicate more accurate and complete vehicle matching results.

\subsection{Evaluation of time alignment}
In this subsection, we evaluate the performance of the proposed time offset optimization algorithm using both  real-world UAV video data (Exp1) and numerical simulation data (Exp2). The algorithm parameters are set to $\theta^{min}=100$, $\theta^{td}=1000$, and $\theta^{scale}=2$.

\begin{figure*}[!ht]
	\centering
	\subfigure[]{\includegraphics[width=0.3\textwidth]{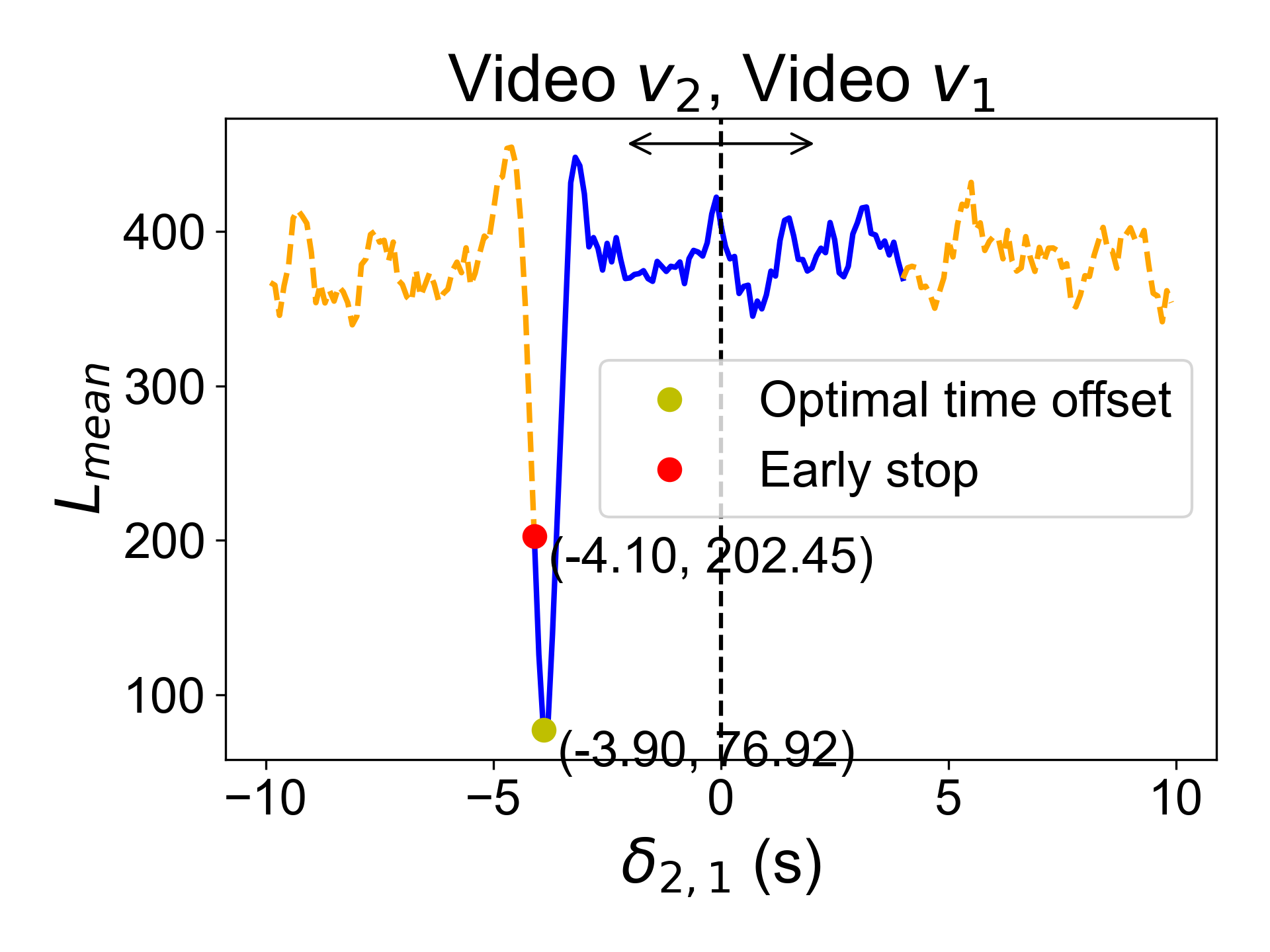}
	\label{fig:time_deviation:a}}
	\subfigure[]{\includegraphics[width=0.3\textwidth]{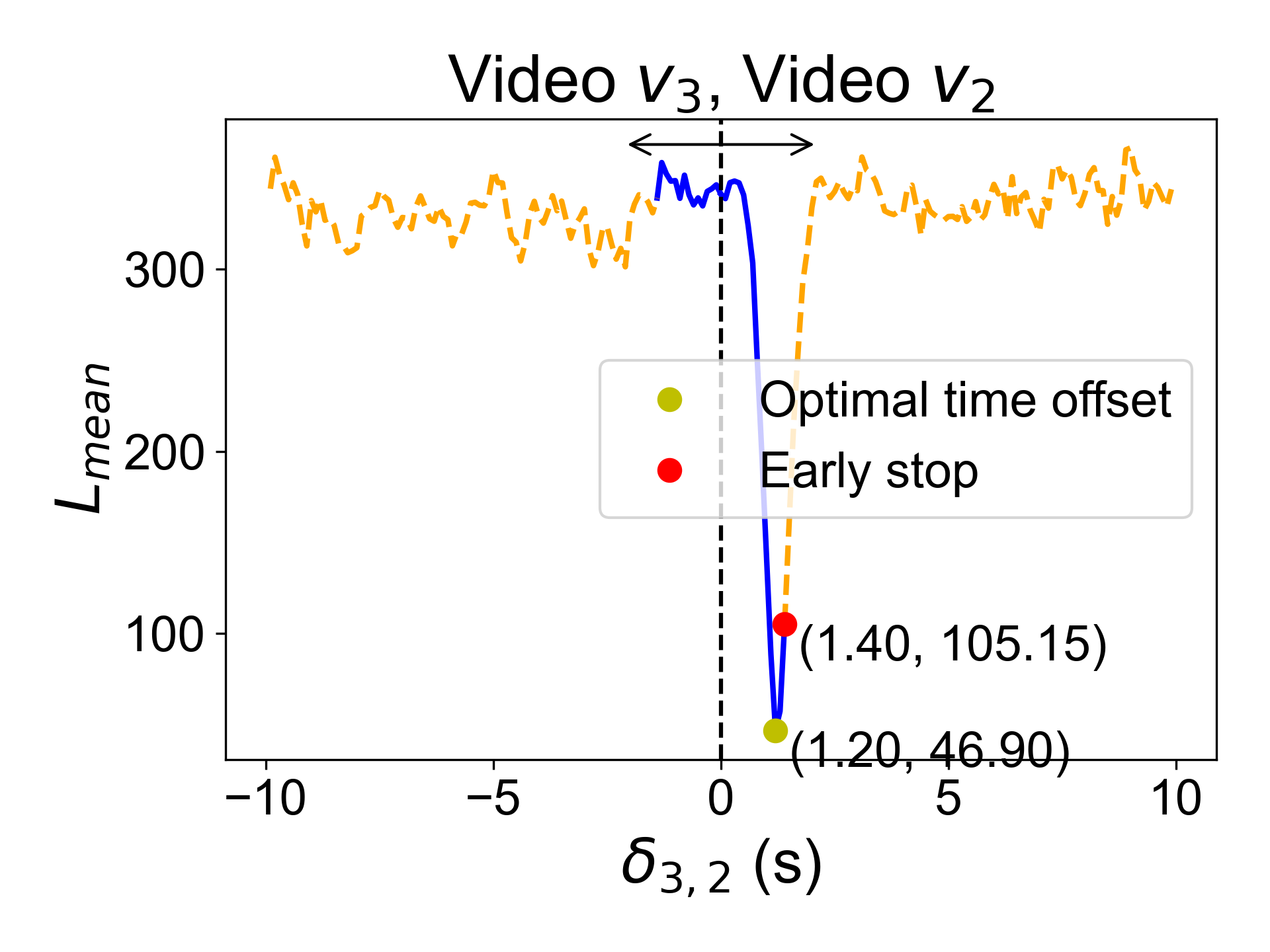}
	\label{fig:time_deviation:b}}
 	\subfigure[]{\includegraphics[width=0.3\textwidth]{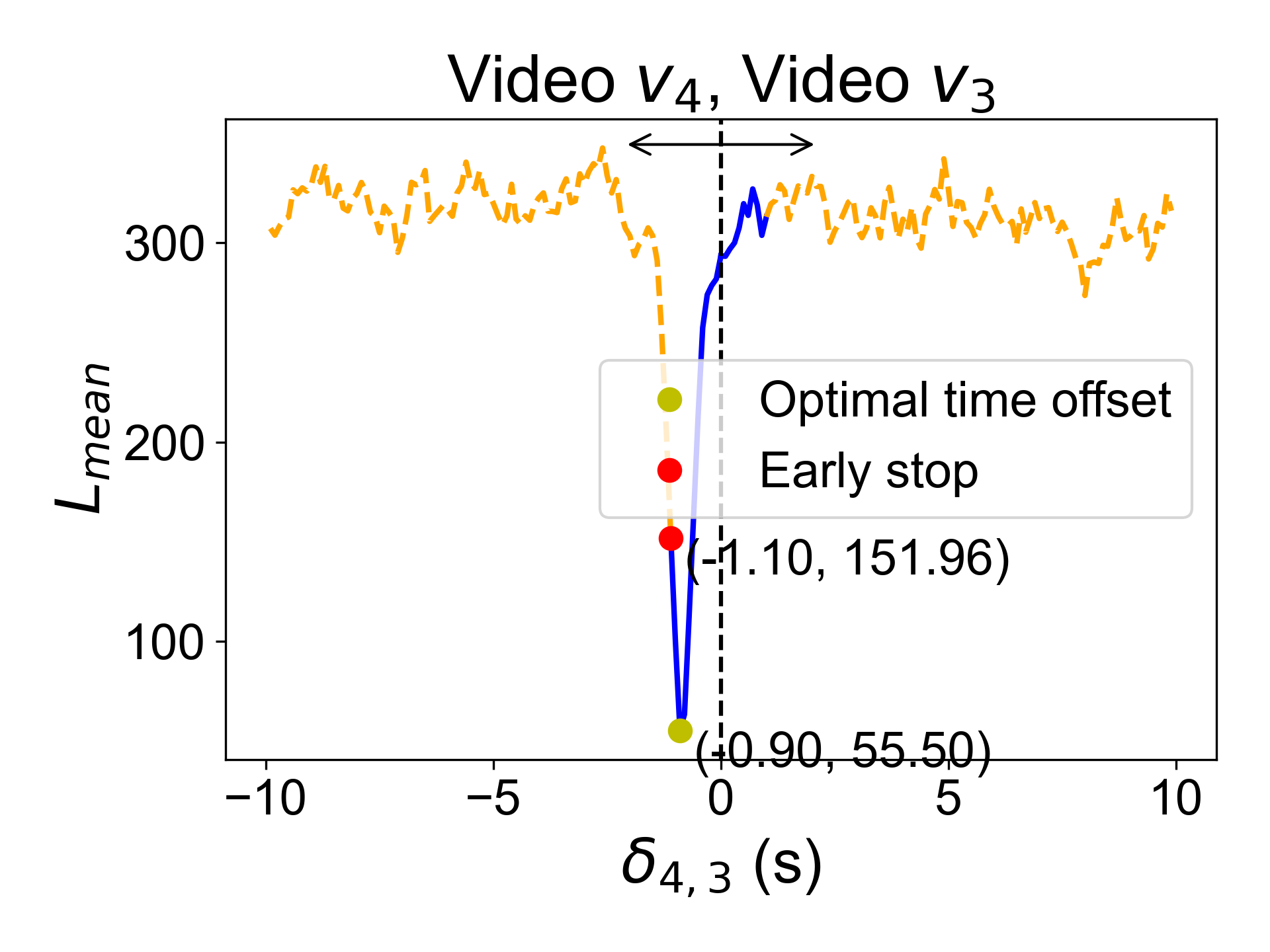}
	\label{fig:time_deviation:c}}
 	\subfigure[]{\includegraphics[width=0.3\textwidth]{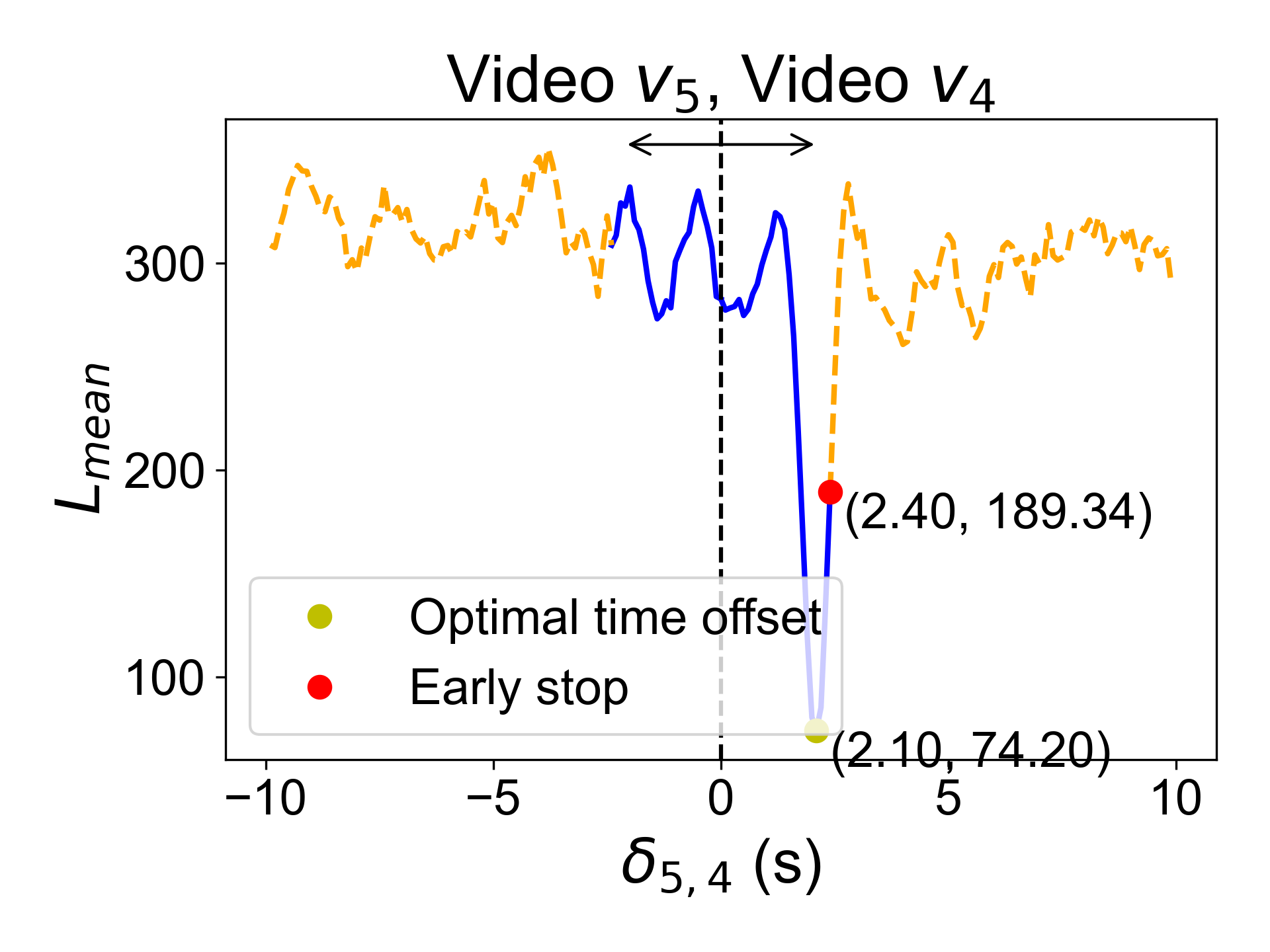}
	\label{fig:time_deviation:d}}
	\subfigure[]{\includegraphics[width=0.3\textwidth]{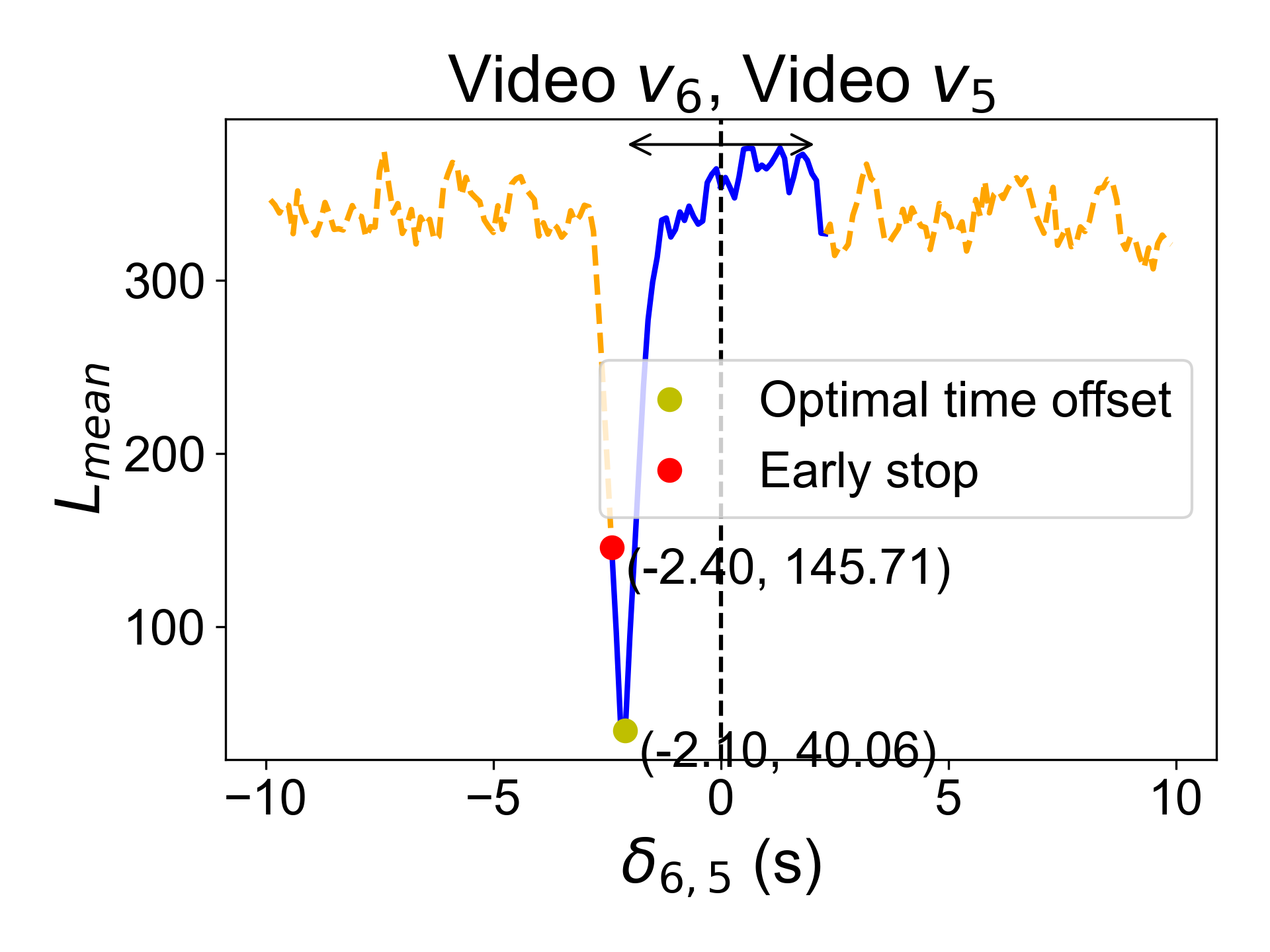}
	\label{fig:time_deviation:e}}
 	\subfigure[]{\includegraphics[width=0.3\textwidth]{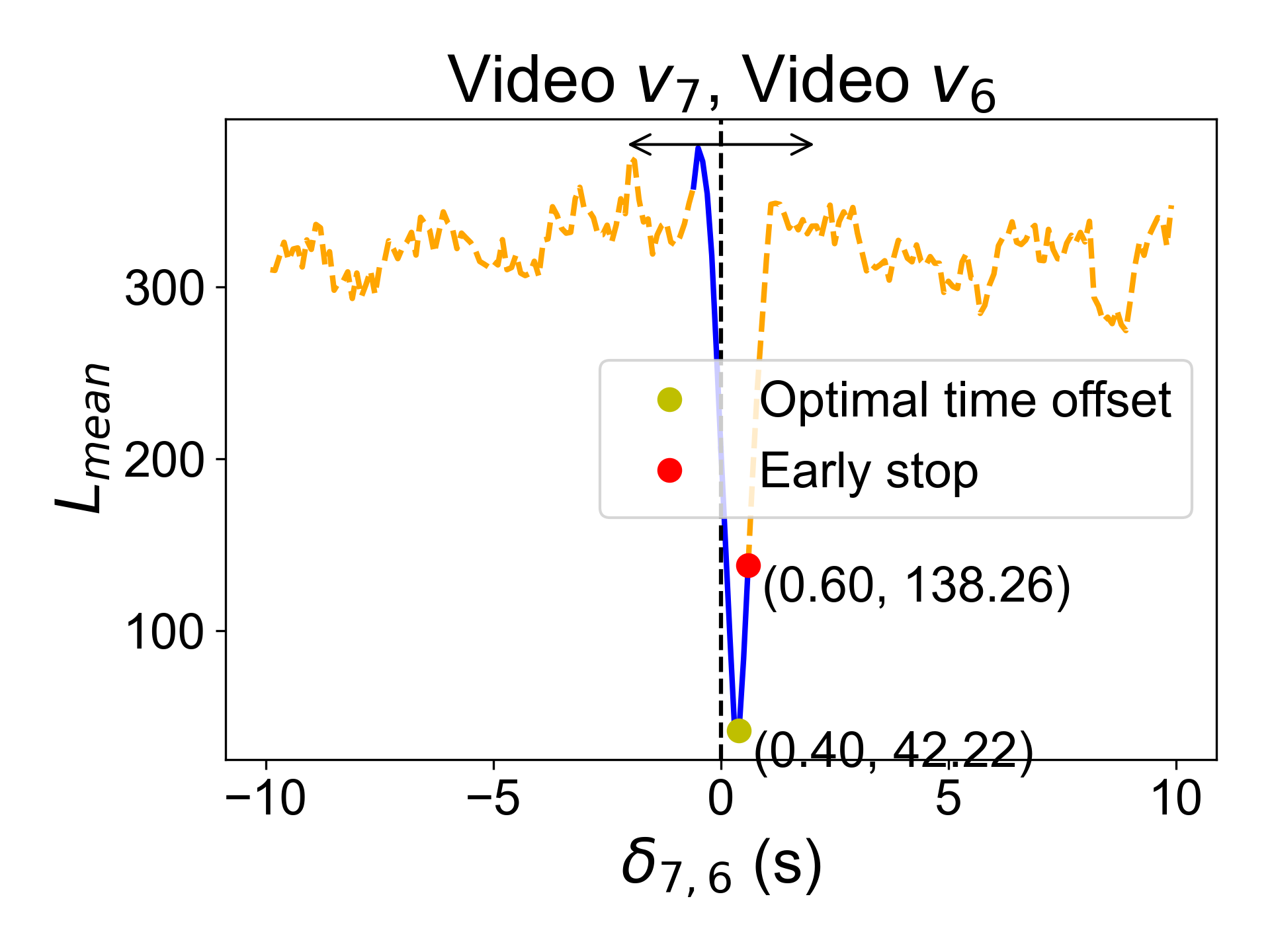}
	\label{fig:time_deviation:f}}

	\caption{Variation of the mean trajectory matching cost defined in Eq.~\eqref{eq:trajectory_difference} for different alignment edges during the search process. The blue curve shows the cost evolution under the bidirectional search, while the orange dashed curve indicates the truncated cost trajectory after the early stopping criterion is triggered.}
    \label{fig:time_deviation}
\end{figure*}

In Exp1, we first analyze the variation of trajectory matching cost under different candidate time offsets using real-world trajectory data in order to examine the identifiability of the optimal time offset in practical scenarios. Fig.~\ref{fig:time_deviation} illustrates the change of trajectory matching cost in the time alignment of the six video pairs captured along the main road. In each subplot, the central double-headed arrow indicates the minimum trajectory matching cost obtained through a bidirectional search starting from zero time offset and exploring both positive and negative directions.
The blue solid line represents the search process with the early stopping strategy, while the orange dashed line corresponds to the search process without employing the early stopping strategy. It is observed that the early stopping strategy significantly reduces the computational workload and improves processing speed.
All subplots exhibit a pronounced global minimum that is significantly lower than neighboring values, indicating that trajectory matching cost is highly sensitive to temporal misalignment. Even small time deviations lead to noticeable increases in trajectory matching cost, which ensures that minimizing the trajectory matching cost can reliably recover the optimal time offset. 
Moreover, the minimum trajectory matching costs across different video pairs fall within a relatively narrow range (approximately 40–80). This consistency suggests that key algorithm parameters, such as $\theta^{min}$ and $\theta^{td}$, do not require scenario-specific tuning, demonstrating the robustness and practical applicability of the proposed method.

\begin{figure}[!ht]
	\centering
	\subfigure[video $v_1$ \& video $v_2$]{\includegraphics[width=0.23\textwidth]{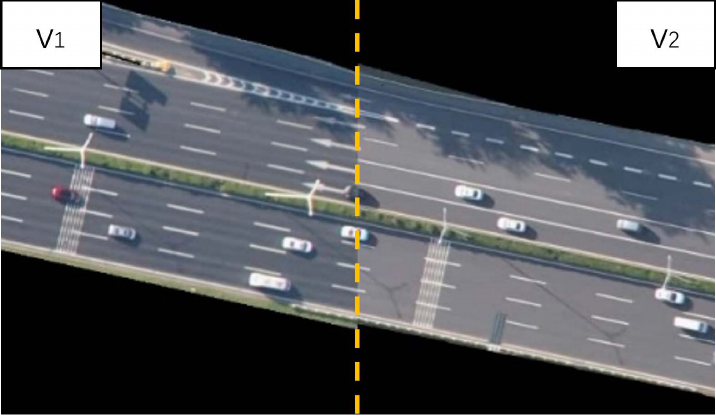}
	\label{fig:eva_time_deviation:a}}
	\subfigure[video $v_2$ \& video $v_3$]{\includegraphics[width=0.23\textwidth]{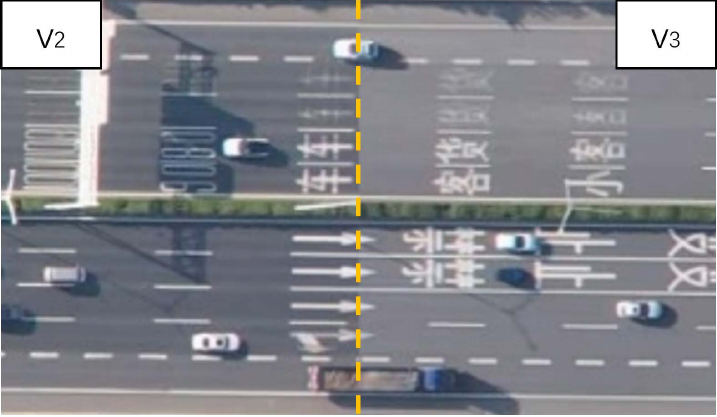}
	\label{fig:eva_time_deviation:b}}
 	\subfigure[video $v_3$ \& video $v_4$]{\includegraphics[width=0.23\textwidth]{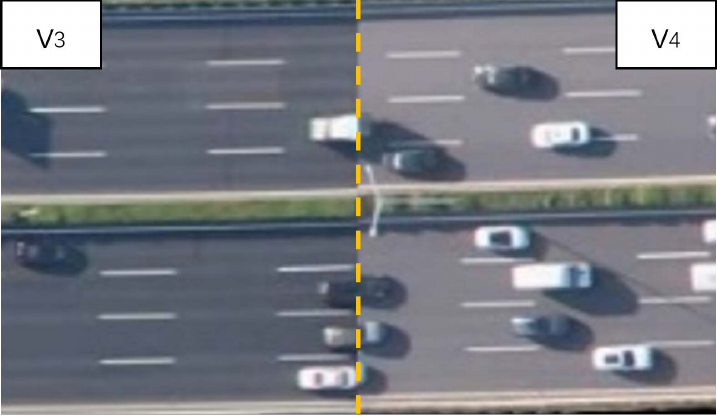}
	\label{fig:eva_time_deviation:c}}
 	\subfigure[video $v_4$ \& video $v_5$]{\includegraphics[width=0.23\textwidth]{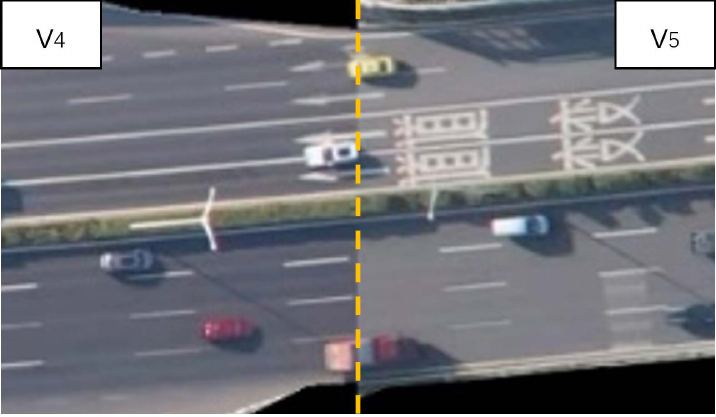}
	\label{fig:eva_time_deviation:d}}

	\caption{The video stitching after time alignment. The yellow dashed lines represent the seams where two videos are stitched together}
    \label{fig:eva_time_deviation}
\end{figure}

Table~\ref{tab:time_deviation} summarizes the time alignment results in Exp1. Here, $\delta_{\text{est}}$ denotes the estimated time offset obtained from the proposed time offset optimization algorithm, while $\delta_{\text{gt}}$ represents the ground-truth time offset determined through manual synchronization.
 As shown in the table, the time offsets vary considerably across different video pairs, underscoring the necessity of explicit time alignment.
The absolute error $|\delta_{\text{est}} - \delta_{\text{gt}}|$ remains small, ranging from 0 to 0.07~s. Given the video frame rate of 30 FPS, this corresponds to a deviation within approximately three frames. These results indicate that the proposed time alignment method achieves high accuracy and reliability.
In practical terms, such a temporal error may translate into a small longitudinal displacement in the stitching region under high-speed freeway conditions. However, its effect is localized to the connection area between adjacent videos. This local offset may induce short-term fluctuations in the estimated speed near the stitching seam, but these fluctuations can be largely mitigated after trajectory smoothing. Moreover, because the temporal misalignment exists only between adjacent videos rather than within a single video, its influence on most subsequent analyses of traffic flow and driving behavior is limited.

To further validate the alignment performance, the estimated time offsets are applied to video stitching for qualitative assessment. Fig.~\ref{fig:eva_time_deviation} presents representative stitched images after time alignment. The yellow dashed lines in these images indicate the stitching seams between adjacent videos. Observing the vehicles at the seam, it is evident that the left and right videos are effectively time-aligned, indicating the high accuracy in the estimated time offsets.



\begin{table}[]
\caption{Time alignment evaluation results}
\label{tab:time_deviation}
\centering
\begin{tabular}{llllll}
\hline
Alignment edge\\ (Video pairs)    & $\delta_\text{est} (s)$ & $\delta_\text{gt} (s)$ & $|\delta_{\text{est}} - \delta_{\text{gt}}| (s)$ \\ \hline
$d_{2,1}$,($v_2,v_1$)  & -3.90  & -3.87 & 0.03     \\
$d_{3,2}$,($v_3,v_2$) & 1.20  & 1.17 & 0.03  \\ 
$d_{4,3}$,($v_4,v_3$) & -0.90  & -0.9 & 0     \\
$d_{5,4}$,($v_5,v_4$) & 2.10  & 2.10 & 0     \\
$d_{6,5}$,($v_6,v_5$) & -2.10  & -2.07 & 0.03     \\
$d_{7,6}$,($v_7,v_6$) & 0.40  & 0.33 & 0.07     \\
$d_{8,7}$,($v_8,v_7$) & -1.50  & -1.43 & 0.07     \\
$d_{9,8}$,($v_9,v_8$) & 1.70  & 1.67 & 0.03     \\
\hline
\end{tabular}
\end{table}

\begin{figure}[!ht]
	\centering
	\includegraphics[width=0.6\textwidth]{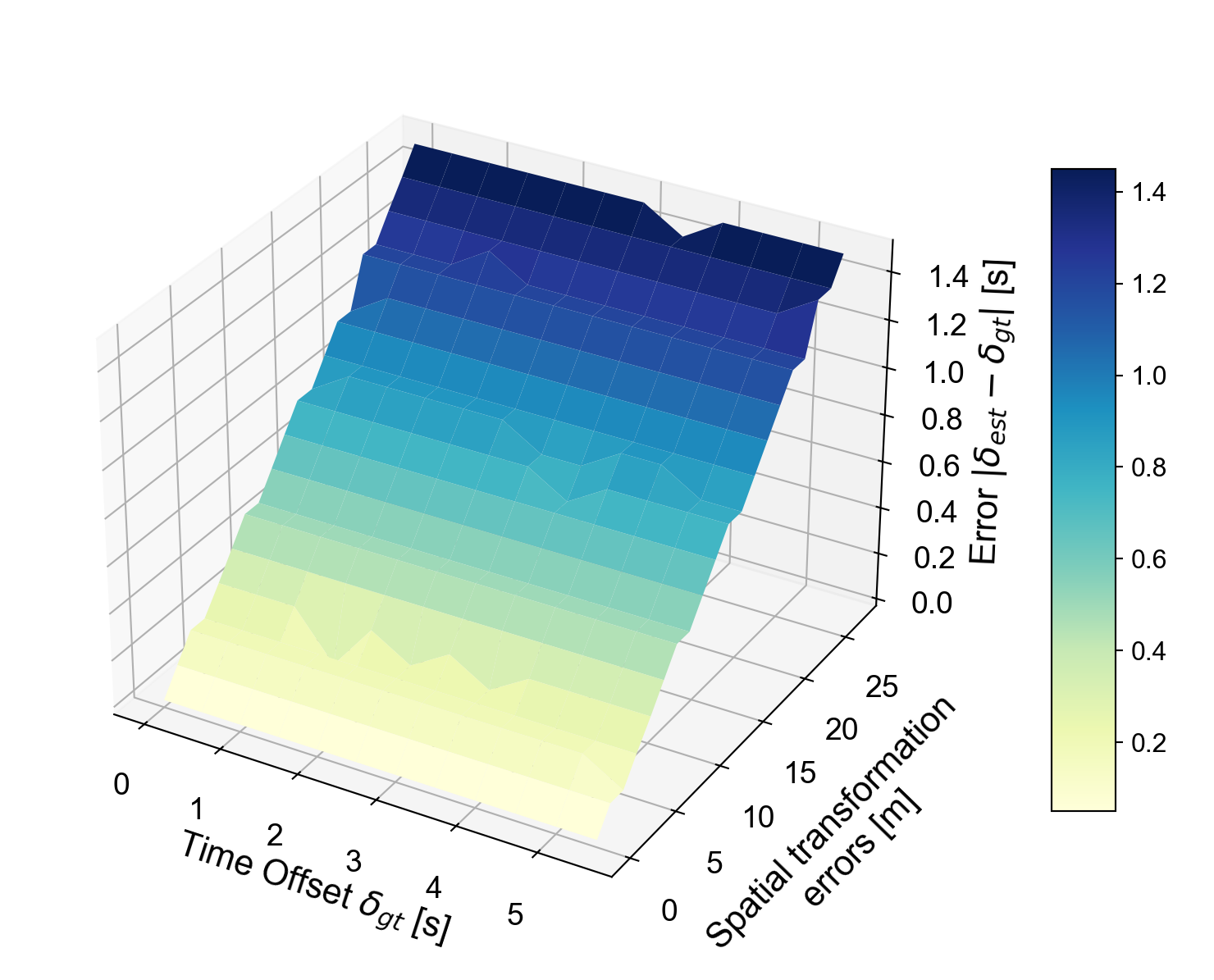}
	\caption{ Time alignment results in Exp2.}\label{fig:exp1_time}
\end{figure}

In Exp2, Fig.~\ref{fig:exp1_time} illustrates the time alignment error under varying ground-truth time offsets and spatial transformation errors. Spatial transformation errors are simulated by adding Gaussian perturbations with mean $\mu$ and a fixed standard deviation of 6 to the ground-truth vehicle positions, thereby mimicking imperfect spatial alignment between overlapping UAV views. As shown in Fig.~\ref{fig:exp1_time}, for a fixed spatial offset, the time alignment error remains nearly constant across different ground-truth time offsets, demonstrating the robustness of the proposed method to varying temporal discrepancies between UAV videos. For a fixed time offset, the alignment error increases approximately linearly with the spatial offset. When the spatial offset approaches zero, the time alignment error is correspondingly close to zero, indicating that accurate spatial transformation is a prerequisite for precise temporal synchronization. These results demonstrate that the proposed time alignment algorithm achieves high accuracy under various time offsets, while also reveal a strong dependency of time alignment accuracy on spatial transformation quality. In this GCVT, high spatial transformation accuracy is achieved through manual annotation of feature points on the first frame of each video, which contributes to improved robustness and precision of the subsequent time alignment process.

\subsection{Evaluation of vehicle matching}
\label{sec:eval_vehicle_matching}
Vehicle matching is a critical component of the trajectory connection framework. Its accuracy depends not only on the matching algorithm itself, but also on the preceding spatial transformation and time alignment stages, which ensure spatial and temporal consistency across videos. Therefore, the evaluation results reflect the overall effectiveness of the proposed GCVT framework. In this subsection, the performance of the vehicle matching module is evaluated using both numerical simulation data and real-world datasets. In addition to quantitative evaluation metrics, the connected vehicle trajectories are visualized to provide qualitative evidence of matching accuracy. The matching threshold parameter in the algorithm is set to $\theta^{matching}=1000$.

In Exp1, three representative alignment edges, corresponding to six videos in total, were selected to evaluate the performance of vehicle matching. The evaluation metrics, including Precision, Recall, and F1-score, as defined in Section~\ref{sec:eval_metric}, were used to quantitatively measure the performance of vehicle matching. We further compared the proposed method with three baseline methods: (1) stitched-video trajectory extraction \citep{chen2020modeling,zheng2024citysim,Chaudhari2025MiTra,Rajput2026SPT}, in which adjacent videos are first stitched and vehicle trajectories are then extracted from the stitched videos. For this method, the time offset obtained from the time alignment step was used for synchronization; (2) appearance-based association, which performs cross-video association using only vehicle appearance embeddings extracted with the DeepSORT feature extractor \citep{Wojke2017DeepSORT}; and (3) HAT \citep{gao2025history}, which dynamically projects ReID features into a sequence-specific discriminative space based on historical trajectory features to enhance multi-object association.

The comparison results are summarized in Table~\ref{tab:vehicle_matching_comparison}. Overall, the proposed GCVT framework consistently outperforms the three baseline methods across the evaluated alignment edges, demonstrating its strong effectiveness and robustness for vehicle matching. Compared with stitched-video trajectory extraction, the proposed method achieves better overall performance, particularly in recall and F1-score on the more challenging edges. Although stitched-video trajectory extraction attains high precision due to prior spatial-temporal alignment, its relatively low recall indicates that many true matches are still missed. This is likely because video stitching is highly sensitive to video stability and alignment quality, and slight camera jitter or local misalignment near stitching boundaries can disrupt vehicle detection and tracking continuity, resulting in interrupted trajectories and missed associations. In addition, its computational cost increases substantially with the number of videos, limiting its scalability for multi-UAV trajectory extraction. The inferior performance of the appearance-based association and HAT baselines further suggests that appearance-based methods are insufficient for reliable vehicle matching in UAV-based multi-video traffic scenes, even when historical trajectory features are incorporated. One likely reason is that aerial videos provide only limited discriminative appearance information, since vehicles are mainly observed from top-down views and many of them appear highly similar. Moreover, the performance of these methods varies substantially across different alignment edges, likely because the degree of appearance distinguishability differs from one edge to another, thereby affecting the robustness of appearance-based matching. Overall, these results verify the effectiveness and robustness of the proposed GCVT framework for multi-UAV vehicle matching.


\begin{table}[!ht]
\caption{Comparison of vehicle matching results}
\label{tab:vehicle_matching_comparison}
\centering
\begin{tabular}{lcccc}
\hline
\textbf{Method} & \textbf{Alignment edge} & \textbf{Precision} & \textbf{Recall} & \textbf{F1-score} \\
\hline
\multirow{3}{*}{Stitched-video trajectory extraction}
& $d_{2,1}$ & 0.999 & 0.630 & 0.773 \\
& $d_{3,2}$ & 0.998 & 0.975 & 0.987 \\
& $d_{4,3}$ & 1.000 & 0.699 & 0.823 \\
\hline
\multirow{3}{*}{Appearance-based association}
& $d_{2,1}$ & 0.449 & 0.451 & 0.450 \\
& $d_{3,2}$ & 0.912 & 0.939 & 0.925 \\
& $d_{4,3}$ & 0.896 & 0.900 & 0.898 \\
\hline
\multirow{3}{*}{HAT}
& $d_{2,1}$ & 0.527 & 0.529 & 0.528 \\
& $d_{3,2}$ & 0.951 & 0.979 & 0.965 \\
& $d_{4,3}$ & 0.960 & 0.965 & 0.963 \\
\hline
\multirow{3}{*}{Proposed GCVT}
& $d_{2,1}$ & 0.999 & 0.982 & 0.990 \\
& $d_{3,2}$ & 0.984 & 0.968 & 0.976 \\
& $d_{4,3}$ & 0.999 & 0.984 & 0.991 \\
\hline
\end{tabular}
\end{table}

To further assess the computational efficiency of the proposed matching framework, we measured the runtime of the Hungarian algorithm, which constitutes the main computational cost in the vehicle matching step, on each alignment edge. The results show that the average runtime is approximately 0.082~s per edge. For the eight alignment edges corresponding to the nine video nodes along the mainline expressway, the total runtime of the Hungarian algorithm is about 0.658~s. These results indicate that the computational burden is low in the current experimental setting.

\begin{figure}[!ht]
	\centering
	\subfigure[video $v_1$ \& video $v_2$]{\includegraphics[width=0.49\textwidth]{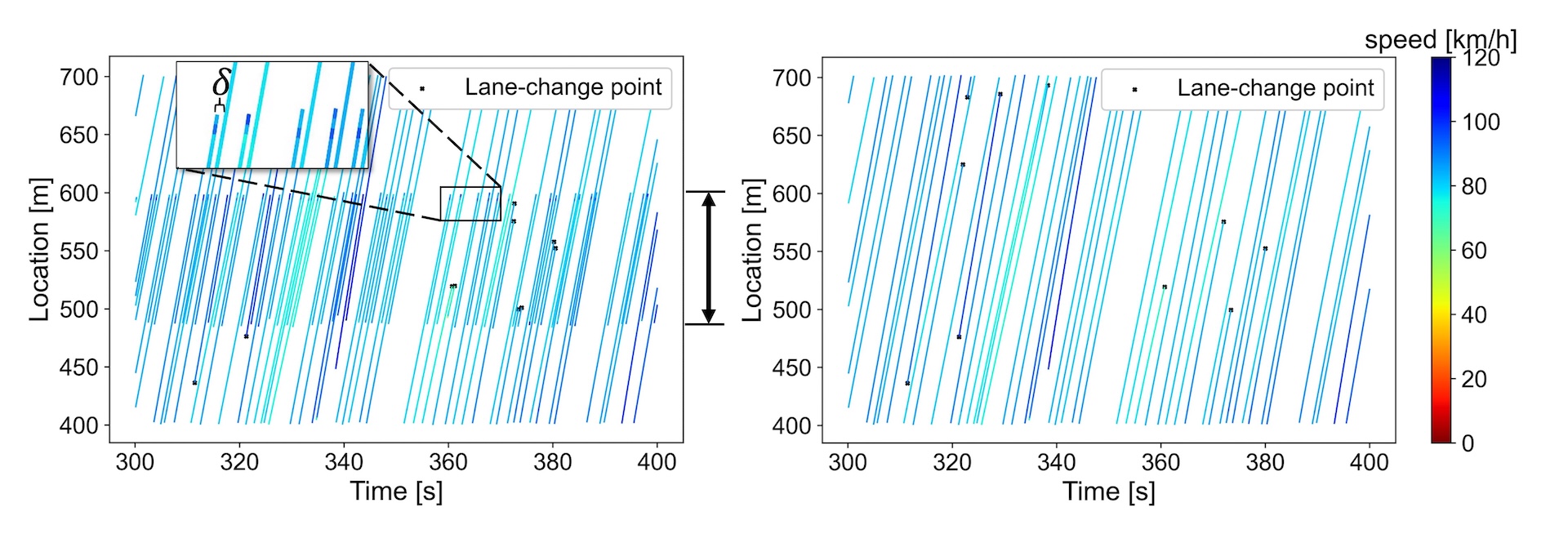}
	\label{fig:eva_match:a}}
	\subfigure[video $v_2$ \& video $v_3$]{\includegraphics[width=0.49\textwidth]{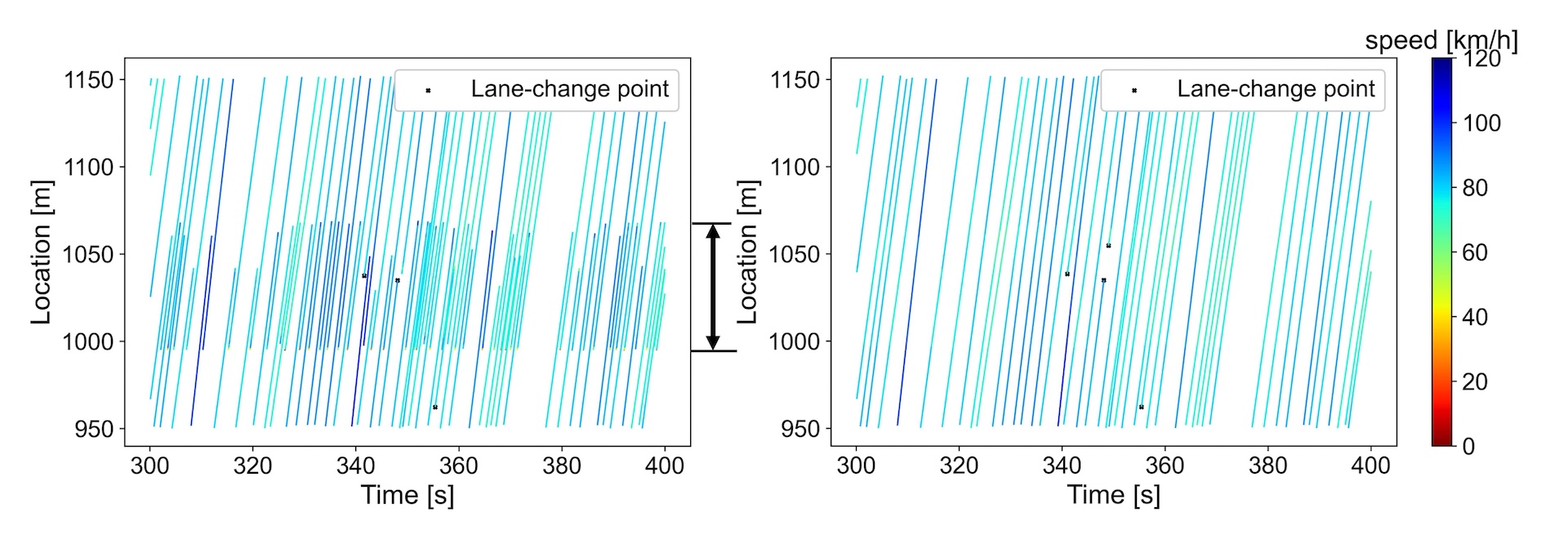}
	\label{fig:eva_match:b}}
 	\subfigure[video $v_3$ \& video $v_4$]{\includegraphics[width=0.49\textwidth]{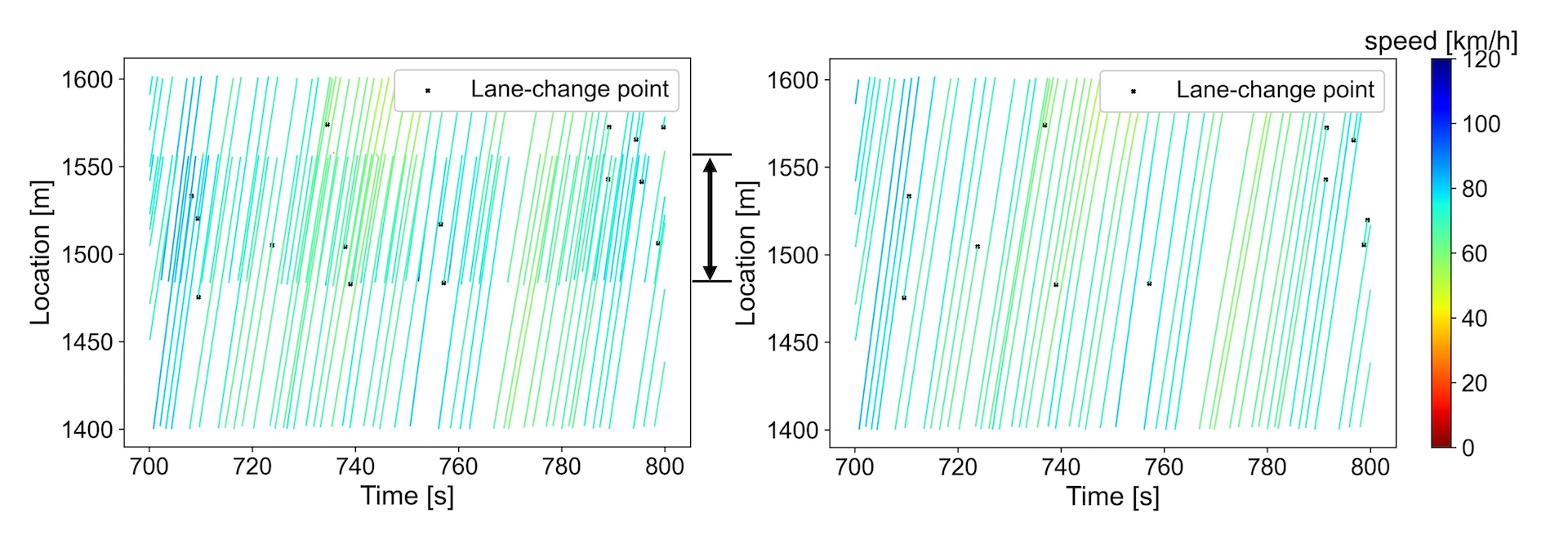}
	\label{fig:eva_match:c}}
 	\subfigure[video $v_4$ \& video $v_5$]{\includegraphics[width=0.49\textwidth]{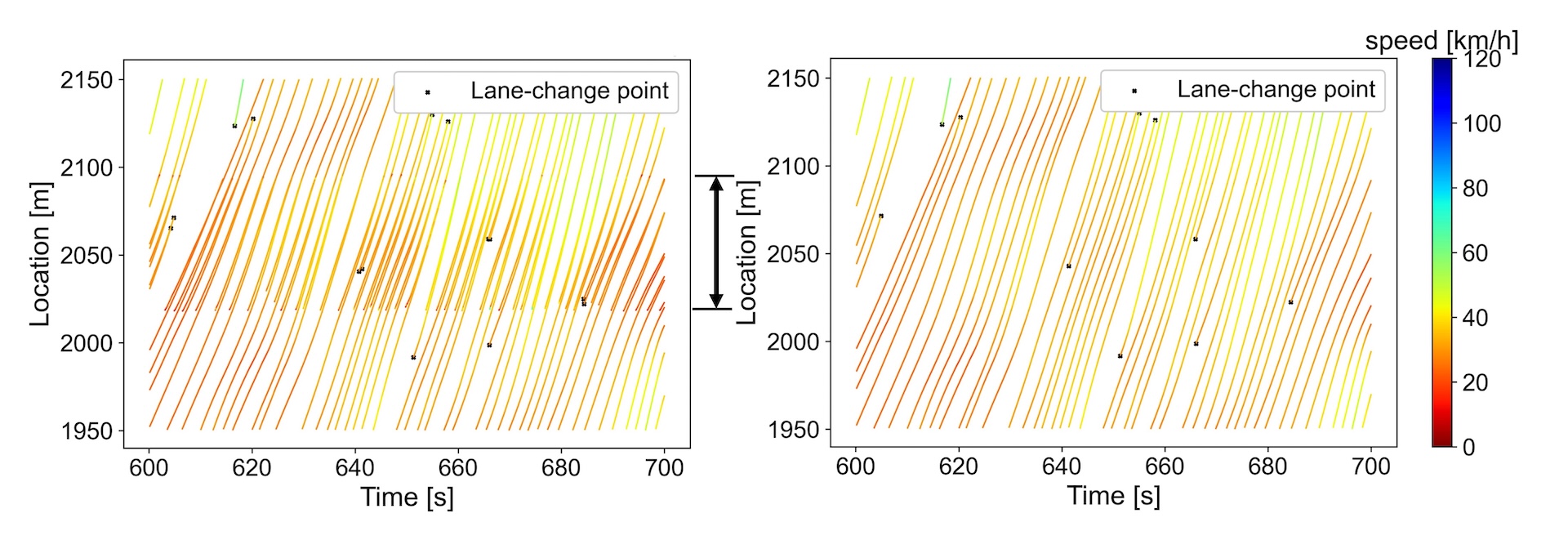}
	\label{fig:eva_match:d}}

	\caption{The original trajectory and connected trajectory.}
    \label{fig:eva_match}
\end{figure}

Fig.~\ref{fig:eva_match} illustrates the comparison between the original trajectory and the connected trajectory. In Fig.~\ref{fig:eva_match:a}-\ref{fig:eva_match:d}, the trajectory on the left side represents the result of spatial transformation alone, while the trajectories on the right side demonstrate the connected trajectory through the application of the proposed GCVT approach. The color of the trajectory indicates the speed of the vehicles. The black double-headed arrow indicates the overlapping area of the two videos. The magnified portion of the trajectory in Fig. \ref{fig:eva_match:a} reveals a distinct misalignment between the upper and lower trajectories. The horizontal distance between these trajectories corresponds to the time offset $\delta$.
 
 From the comparison of the trajectories before and after connection, it is apparent that the proposed method can effectively eliminate the sudden changes in trajectory caused by time misalignment, resulting in smoother trajectories.
 The colors of the trajectories in Fig.\ref{fig:eva_match:a} to \ref{fig:eva_match:d} depict the variation in speed, encompassing both free-flowing and congested traffic. This observation highlights the method's applicability for stitching under different traffic flow conditions.
 Each subplot has lane-changing events within the video overlapping area, which suggests that the GCVT approach can also achieve successful  trajectory connections for vehicles changing lanes in the overlap area, without relying on car-following behavior. The approach's capability to accurately connect trajectories in such complex traffic situations highlights its robustness and adaptability to various driving behaviors.

\begin{figure*}[!ht]
	\centering
	\includegraphics[width=0.8\textwidth]{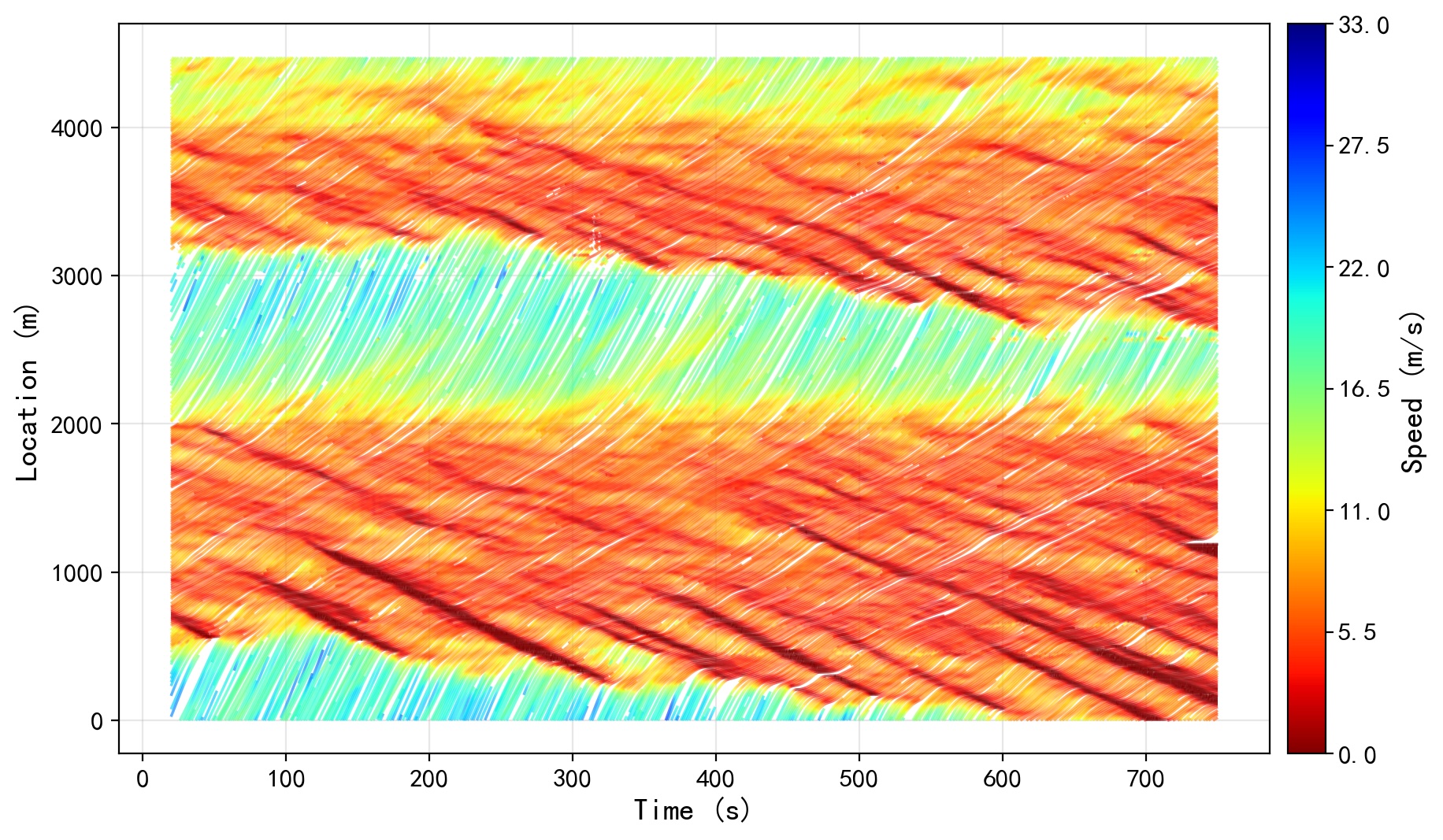}
	\caption{Space–time diagram of vehicle trajectories after connection, covering videos $v_1$–$v_9$.}
    \label{fig:full_trajectory}
\end{figure*}

\begin{figure*}[!ht]
	\centering
	\includegraphics[width=0.9\textwidth]{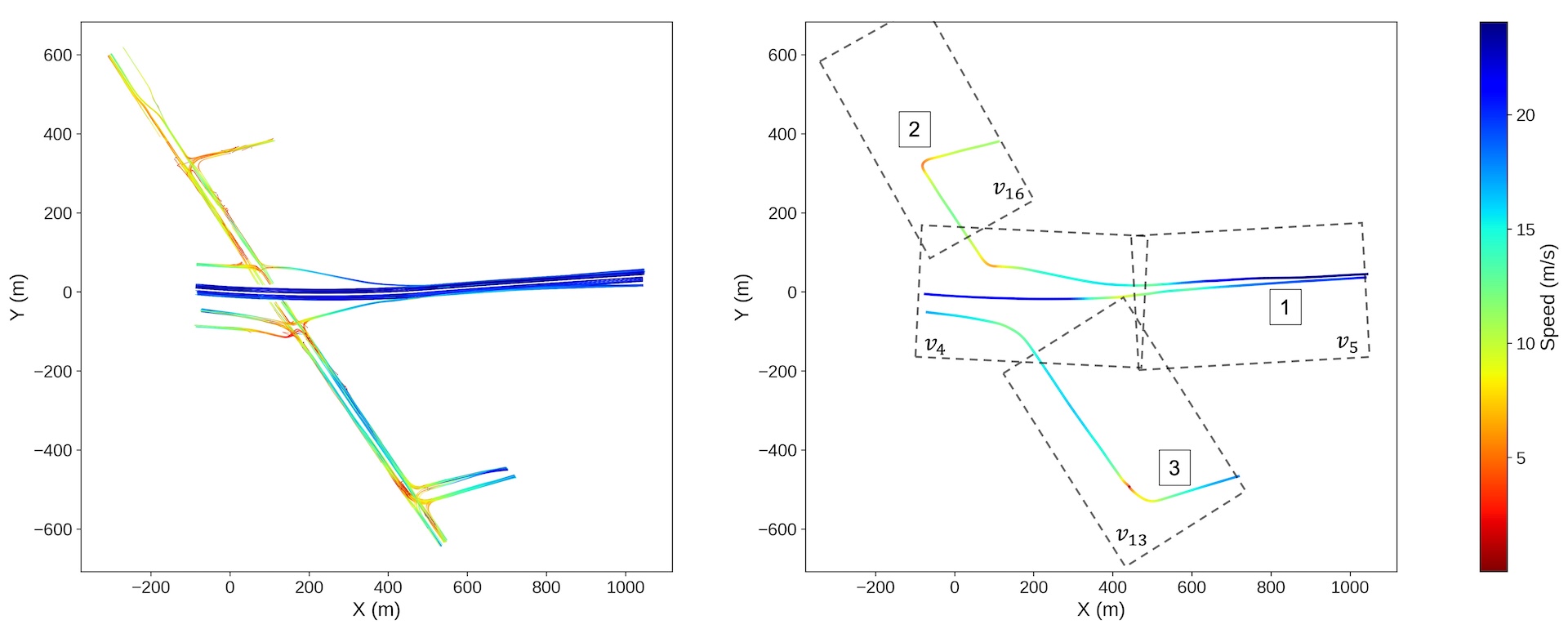}
	\caption{Spatial trajectories illustrating connections across videos $v_4$, $v_5$, $v_{13}$, and $v_{16}$.}
    \label{fig:trajectory_2d}
\end{figure*}

After connecting vehicles across each pair of videos, the complete set of vehicle trajectories captured by the UAV swarm can be obtained, as shown in Fig.\ref{fig:full_trajectory}. The figure illustrates the vehicle trajectories on lane 1 of the mainline segment, covering from $v_1$ to $v_9$ (as shown in Fig.\ref{fig:video_graph_h}), where different colors represent vehicle speeds. The trajectories extend over 4,000 meters, and a fixed bottleneck can be observed around the 2,000-meter mark, corresponding to a merging area.
Beyond expressway trajectories, the proposed GCVT method is capable of connecting vehicle trajectories recorded by irregularly distributed UAVs, enabling the reconstruction of trajectories across complex network structures involving expressways and urban intersections. Fig.~\ref{fig:trajectory_2d} presents the spatial trajectories obtained by stitching videos $v_4$, $v_5$, $v_{13}$, and $v_{16}$. The left figure shows all stitched trajectories during the first three minutes, while the right figure highlights three representative trajectories. The dashed boxes indicate the spatial coverage of each video, and the rectangular labels denote vehicle IDs.
As illustrated, video $v_4$ is connected not only to the expressway segment captured in $v_5$ but also to two urban intersections recorded in $v_{13}$ and $v_{16}$. Among the representative trajectories, Vehicle~1 travels along the expressway, whereas Vehicles~2 and~3 travel from the expressway into urban roads. These results demonstrate that the proposed method can effectively stitch trajectories captured by irregularly distributed UAVs, enabling the reconstruction of continuous vehicle trajectories across heterogeneous road types.
 
\begin{figure}[!ht]
	\centering
	\subfigure[]{
		\includegraphics[width=0.4\textwidth]{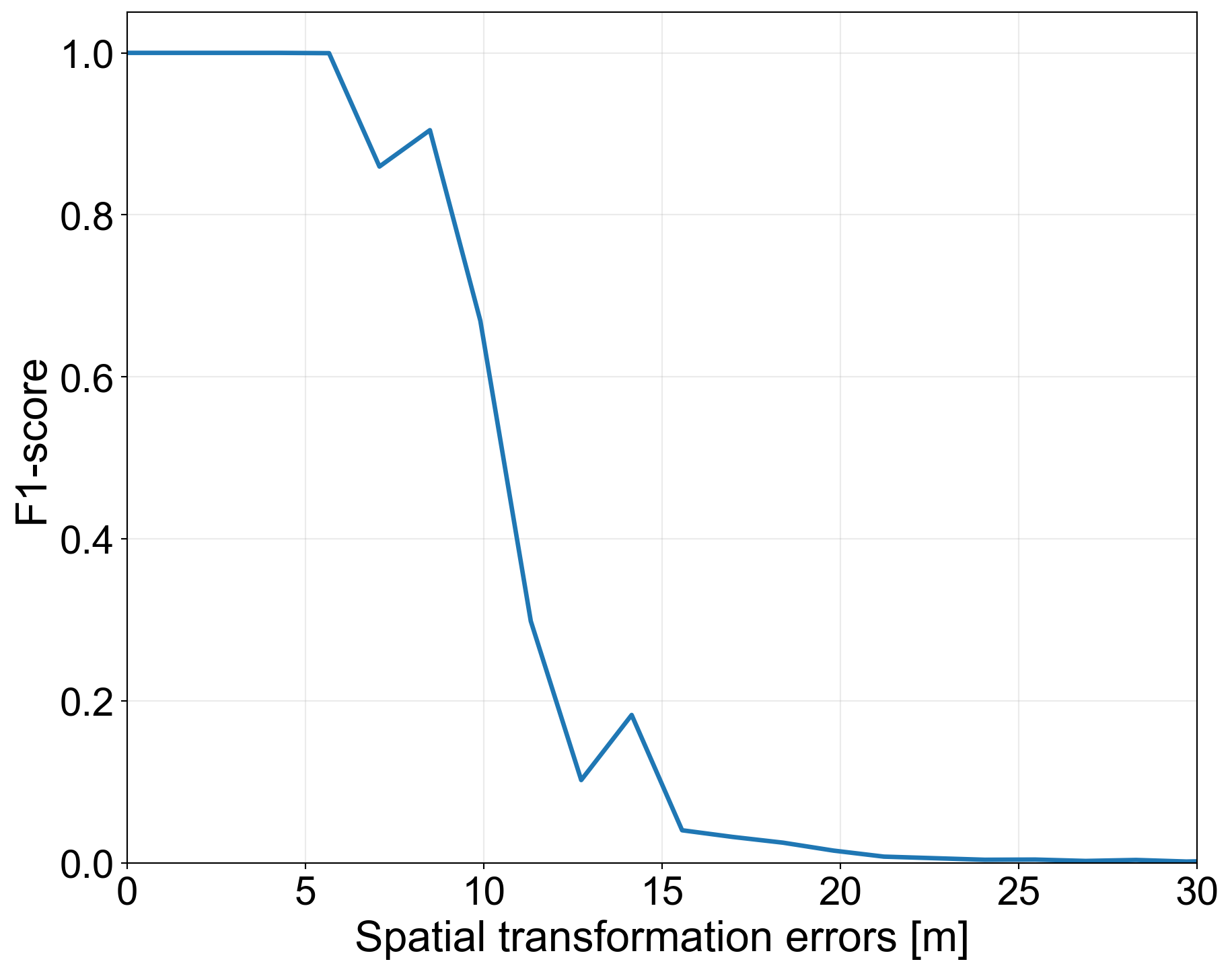}
		\label{fig:exp2_matching:a}
	}
	\subfigure[]{
		\includegraphics[width=0.4\textwidth]{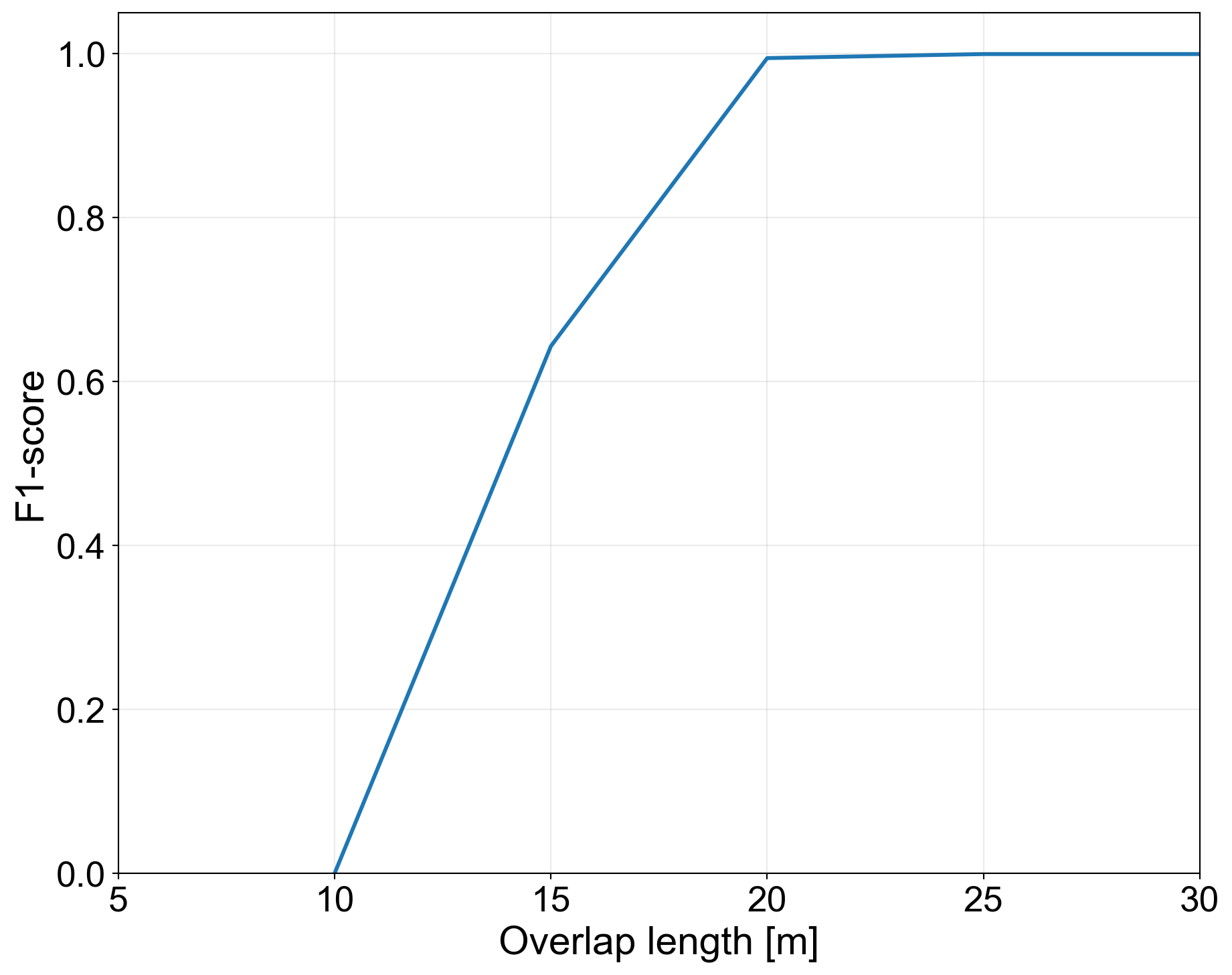}
		\label{fig:exp2_matching:b}
	}
	\caption{Vehicle matching results in Exp2: (a) F1-score under different spatial transformation errors; (b) F1-score under different overlap lengths.}
	\label{fig:exp2_matching}
\end{figure}


In the time alignment analysis of Exp2, similar time alignment errors were observed across different ground-truth time offsets under the same spatial offset. Therefore, the vehicle matching analysis in Exp2 focuses on two practical factors that directly influence vehicle matching: spatial transformation error and overlap length. Spatial transformation error characterizes the geometric misalignment caused by imperfect spatial calibration, whereas overlap length denotes the length of the road segment jointly captured by two adjacent videos, which determines the amount of shared trajectory information available for matching.

Fig.\ref{fig:exp2_matching}(a) shows the F1-score under different spatial transformation errors, while Fig.\ref{fig:exp2_matching}(b) shows the F1-score under different overlap lengths. As shown in Fig.\ref{fig:exp2_matching}(a), the F1-score remains close to 1.0 when the spatial transformation error is below approximately 6~m. Since a 6~m error is approximately equivalent to the width of two traffic lanes, this result demonstrates that the proposed method remains highly robust even under considerable geometric misalignment. As shown in Fig.\ref{fig:exp2_matching}(b), the F1-score increases substantially with the overlap length, rising from nearly zero at 10~m to about 0.64 at 15~m, and reaching almost 1.0 at 20~m and above. This result indicates that a sufficient overlap length is essential for accurate and stable vehicle matching, while also showing that the proposed method does not require an excessively long overlapping region in practical deployment.

\subsection{Ablation study}
To quantify the contribution of the key modules in the proposed framework, we conducted an ablation study on the vehicle matching performance using the SWIFTraj dataset, by removing or simplifying one component at a time. The evaluated variants include: (1) the proposed GCVT framework, which includes both the time alignment module and the original two-step Hungarian matching design; (2) a variant without time alignment, in which vehicle matching is performed directly after spatial transformation without correcting time offsets between adjacent videos; (3) a variant without the lane-related term in the trajectory matching cost \(L_{i,j}\), where the trajectory matching cost is computed using only pixel-coordinate positions; and (4) a variant with single-step Hungarian matching, in which the Hungarian algorithm is applied only once without the second-step matching used in the proposed framework. Table~\ref{tab:ablation} summarizes the results.

As shown in Table~\ref{tab:ablation}, the proposed GCVT framework achieves the best overall performance, with an F1-score of 0.973. Removing the time alignment module causes the most significant degradation, reducing the F1-score to 0.890, which confirms the importance of correcting time offsets between adjacent videos. This result also suggests that relying only on the original clocks of the UAVs is insufficient to achieve reliable vehicle matching. Removing the lane-related term in \(L_{i,j}\) and replacing the original two-step Hungarian matching with a single-step strategy also reduce the performance, demonstrating that both components contribute to the final trajectory connection accuracy. Overall, the ablation results verify the effectiveness of the main modules in the proposed framework.

\begin{table}[!ht]
\caption{Ablation study results}
\label{tab:ablation}
\centering
\begin{tabular}{lccc}
\hline
\textbf{Variant} & \textbf{Precision} & \textbf{Recall} & \textbf{F1-score} \\
\hline
Proposed GCVT & 0.976 & 0.971 & 0.973 \\
w/o time alignment & 0.880 & 0.901 & 0.890 \\
w/o lane term in \(L_{i,j}\) & 0.957 & 0.983 & 0.970 \\
Single-step Hungarian matching & 0.931 & 0.957 & 0.944 \\
\hline
\end{tabular}
\end{table}

\section{Conclusion}
\label{sec:conclusion}
To achieve the automatic time alignment of multi-UAV videos and the connection of vehicle trajectories, this paper proposes a novel graph-based vehicle trajectory connection approach. 
The framework models UAV layouts as a graph, where each video corresponds to a node and overlapping regions define edges, allowing flexible handling of irregular UAV deployments.
A major strength of our approach lies in its ability to perform automatic time alignment by minimizing the trajectory matching cost. This ensures accurate temporal synchronization of vehicle trajectories across videos. 

Both real-world and simulation experiments were conducted to validate the proposed framework, with a focus on time alignment and vehicle matching. In the real-world experiment involving 16 UAVs, the absolute time alignment error was within three video frames. Quantitative evaluation based on Precision, Recall, and F1-score indicates near-perfect vehicle matching performance. The reconstructed trajectories further demonstrate the capability of the framework to handle diverse traffic states, lane-changing scenarios, large spatial coverage, and irregular UAV layouts.

Simulation experiments with varying time offsets and spatial transformation errors further reveal that time alignment remains stable under different time offsets when spatial transformation is accurate. Vehicle matching performance gradually decreases as spatial error increases and drops sharply beyond a certain threshold. These findings confirm the robustness of the proposed approach to temporal discrepancies and its tolerance to spatial transformation errors within a reasonable range.


For future research, the proposed GCVT framework can be extended to broader multi-surveillance scenarios. Since the method operates directly on trajectory data without restricting its source, it can be applied to integrate trajectories from multiple road monitoring systems to generate continuous vehicle trajectories.
Several promising directions can be pursued to enhance the proposed GCVT approach. First, the current trajectory matching cost calculation for vehicle matching is based solely on pixel coordinates and lane ID. Incorporating additional vehicle attributes, such as vehicle category and appearance features, may further improve matching accuracy and robustness. 
Furthermore, the current method of manually marking feature points is employed to ensure the accuracy of spatial transformation. Second, spatial transformation currently depends on manual annotation of feature points to ensure high precision. Future work may explore hybrid strategies that combine automatic feature detection with manual verification, thereby improving efficiency while maintaining spatial accuracy.

\section*{Code}
\label{sec:code}
The proposed GCVT approach  has been incorporated into our vehicle trajectory extraction framework, OpenVTER. The source code is available at https://github.com/XinkaiJi/OpenVTER, and the dataset can be accessed at https://swiftraj.com.

\section*{Acknowledgment}

This research is supported by the National Natural Science Foundation of China (No.52525204, No.52232012).

\bibliographystyle{elsarticle-harv} 
\bibliography{main}

\end{document}